\documentclass[aps,preprintnumbers,nofootinbib,floats,floatfix]{revtex4-2}
\usepackage{amsfonts}
\usepackage{amsmath}
\usepackage{graphicx}
\usepackage{amssymb}
\usepackage{amsxtra}
\usepackage{mathrsfs}  
\usepackage{amstext}
\usepackage{url}
\usepackage{orcidlink}
\usepackage{endnotes}
\usepackage{xcolor}
\usepackage{hyperref}
\hypersetup{hidelinks}
\begin{document}
\title{Reducible first-class constraints, gauge symmetry, and the degrees of freedom of three-dimensional gravity coupled to topological matter}
\author{Omar Rodr\'iguez-Tzompantzi \orcidlink{0000-0002-6307-7203}} \email{omar.tzompantzi@unison.mx}
\affiliation{Departamento de Investigaci\'on en F\'isica, Universidad de Sonora,
Apartado Postal 5-088, C.P. 83000, Hermosillo, Sonora, M\'exico.}
\begin{abstract}
We develop a systematic Hamiltonian formulation for a gravitating topological matter system in three-dimensional spacetime, coupling a scalar matter field and a two-form gauge field to first-order Hilbert--Palatini gravity. We perform the Dirac--Bergmann analysis, obtaining the full structure of the constraints, classifying them into first- and second-class ones, and computing their Poisson bracket algebra. Furthermore, we write down the explicit expression for the Hamiltonian generator of gauge symmetries on the full set of canonical variables, containing the exact number of gauge parameters. A mapping of these parameters then shows that the gauge transformations reproduce on-shell the spacetime diffeomorphism and local Poincar\'e symmetries. The symmetry structure of the coupled model is thereby established on a Cauchy surface without boundary. Our canonical analysis further shows that the first-class constraints obey three reducibility conditions. We trace them back to an off-shell identity of Ricci type and to the redundancy of the gauge parameters carried by the two-form field. In the extended phase-space they close only modulo the second-class sector. Once they are taken into account, the model carries no local degrees of freedom. The fundamental symplectic structure on the reduced phase-space is fixed by the explicit computation of the Dirac brackets. Both these and the Hamiltonian gauge generator are obtained in the extended phase-space, with no canonical variable eliminated beforehand. Finally, we show that on a surface with boundary the improved generator built from the reducible parameters collapses to a single boundary integral, which vanishes identically.
\end{abstract}
% \date{\today}
%\pacs{}
\preprint{}
\maketitle
\section{Introduction}
Lower-dimensional field-theoretic models provide a powerful framework for gaining deeper insight into the fundamental interactions of Nature. For instance, unlike four-dimensional ($4$d) general relativity, $3$d pure gravity—described by the Einstein--Hilbert action—exhibits no local propagating physical degrees of freedom (DoF) and, consequently, features neither Newtonian forces nor gravitational waves \cite{Leutwyler, Deser, Deser2}. Furthermore, with or without a cosmological constant, the theory admits a first-order formulation as a gauge theory, governed by the Hilbert--Palatini action \cite{Montesinos,Duque}.\footnote{We follow the convention that is standard in the canonical literature: Hilbert--Palatini names the action functional, while Einstein--Cartan names the theory, and the Riemann--Cartan geometry, that this action describes \cite{Duque,Hohm}. The theory is also known as Einstein--Cartan--Sciama--Kibble, after the authors who worked out how the connection couples to spin \cite{Utiyama,Kibble}.} The action so named is the functional $S[e,A]$ of a vielbein---the dreibein, in three dimensions---and an independent Lorentz connection, and not the metric-affine functional $S[g,\Gamma]$; the two are classically equivalent if, and only if, $e$ is invertible \cite{Sadovski}. In this formulation both local Lorentz and diffeomorphism invariance are manifest \cite{Peldan,Witten}. At the same time, $S[e,A]$ is not itself independent of the spacetime metric, since it is built from the soldering form $e$. 

What is metric-independent is the Chern--Simons description in terms of the connection alone. The equivalence holds provided the spacetime dimension is odd, the connection is flat and $e$ is invertible \cite{Witten, Achuarro, Romano, Witten2}. The same condition applies to the Chern--Simons-like actions of three-dimensional massive gravity, whose equivalence to the usual ones can be established only for an invertible dreibein \cite{Hohm}. This equivalence survives perturbative quantum corrections, but there are reasons to expect the non-perturbative regime to be dominated by singular configurations of $e$, in which case the topological description need not persist \cite{Witten2}; in the metric formulation one integrates only over invertible dreibeins, whereas the gauge formulation admits the singular ones as well, and it is at this point that the two descriptions differ \cite{ChenTe}. In the presence of torsion the correspondence is further restricted: topological three-dimensional Riemann--Cartan gravity of Mielke--Baekler type admits an $SL(2,\mathbb{R})\times SL(2,\mathbb{R})$ Chern--Simons form provided one stays in its anti-de Sitter sector, defined by a negative effective cosmological constant \cite{Vasilic}. 

The absence of local degrees of freedom and the gauge-theoretic formulation significantly simplify the analysis of $3$d gravity relative to its $4$d counterpart, resulting in an exactly solvable theory that can be quantized through various approaches \cite{Carlip2, Carlip, Noui, Kim}.  In addition, since $3$d and $4$d gravity share a common foundational structure—characterized by diffeomorphism invariance, the absence of background time, and constraint-encoded dynamics—$3$d gravity serves as a robust toy model for exploring the complexities of the gravitational field itself and several issues in quantum gravity \cite{Ashtekar}. These include the nature of time in quantum gravity \cite{Measuring}, the meaning of diffeomorphism-invariant observables \cite{Observables}, and foundational questions such as whether gravitational interactions may have a finite range \cite{Boulware}.

Going further, a consistent quantum theory of gravity is expected to bear on several long-standing problems. Among the ones that have been posed in the literature are the resolution of spacetime singularities, the late-time accelerated expansion of the Universe, black hole evaporation and the information paradox, and the gravitational field sourced by matter in spatial superposition \cite{Superposition, Bose}. Since all these questions involve matter in one way or another, it is natural to ask how far pure gravity models can take us. Whether matter is essential in quantum gravity is not known, and the results available are model-dependent; this is in itself a reason to study matter couplings that remain tractable. Three dimensions offer a concrete illustration of what is at issue: once gravity is integrated out, matter coupled to it acquires a momentum space with non-trivial geometry, and its kinematics is governed by deformed, group-valued symmetries \cite{Arzano}. Bearing this in mind, we must search for interacting theories that consistently couple the structural versatility of $3$d gravity to the dynamics of matter fields. In particular, what canonical approaches to quantum and to modified gravity with matter demand is that the Hamiltonian formulation of such theories be worked out in full, gauge content included \cite{Duque}.

In recent years, gravity in three dimensions has also proven highly useful for describing the mechanisms behind exotic physical phenomena realized in certain condensed matter systems (see, e.g., Refs. \cite{Haldane,Igor}). This kind of application links gravitation to condensed matter physics, and suggests that each of the two fields has something to gain from the other. However, coupling matter to gravitation in three dimensions in the traditional way—either through the metric in the second-order formalism or through the dreibein field in the first-order formulation—generally spoils the topological properties of pure Einstein theory and possibly introduces local dynamics. As a result, the quantization process faces many of the same obstacles as its $4$d counterpart. A remarkable exception lies in certain special classes of topological field theories that couple matter to gravity through the connection in the first-order formalism (rather than the dreibein), while still preserving the solvability characteristic of the pure gravity theory \cite{Horowitz,SCarlip,Gegenberg1,Burwick,Marcela}. 

We focus on one such exceptional class: three-dimensional first-order gravity coupled with topological matter. Topological matter is studied at once in several fields, from electromagnetism and high-energy physics to atomic-molecular optics and condensed-matter physics, on account of the unusual electromagnetic and gravitational effects it often exhibits \cite{Wen,Hasan}. Theoretically, these systems involve not only point-like quasi-particles—such as fermions, bosons, and anyons—but also extended objects like quasi-strings and quasi-membranes \cite{FHE}. These extended objects can be conventionally described by two-form, that is, rank-$2$ antisymmetric, gauge fields through effective gauge potentials with topological terms \cite{Goldman}. In our specific setup, the topological matter sector consists of a scalar field and a two-form field—also known as a Kalb--Ramond field\footnote{Antisymmetric tensor gauge fields (two-forms) have proven highly relevant in diverse areas such as string and membrane theory \cite{Kalb}, gravity \cite{Speziale}, black hole physics \cite{Asrat}, spin foams \cite{Baez}, and the topological mass generation mechanism \cite{Medeiros, Aurilia}. In modern condensed matter physics, these fields appear naturally in interacting topological insulators \cite{Pachos, Chan} and characterize a wider class of strongly interacting spin systems known as spin liquids \cite{Pretko, Kristian} in the low-energy regime.}, whose coupling to gravity preserves the solvability of the pure $3$d theory. The interest in these systems is not restricted to the formal side. There are already concrete proposals for simulating three-dimensional first-order gravity coupled to matter in the laboratory. The Dirac fermions arise in the low-energy limit of a honeycomb lattice, and the gravitational fluctuations are carried by bosonic modes placed on its links \cite{Salgado}.

While a spacetime-dependent effective action can describe both topological and dynamical aspects of topological matter, a phase-space treatment has proven convenient and efficient in the Chern--Simons descriptions of such systems \cite{Palumbo,Tomoya}. One qualification should be added before carrying this over to gravity: a canonical analysis is available only on globally hyperbolic sectors of the space of solutions, so it cannot be said to capture every aspect of a gravitational theory. Furthermore, although the lack of local DoF is well understood in pure $3$d gravity, coupling to matter—even of a topological nature—requires careful verification that no local excitations emerge from the interaction terms. Given that dynamical systems of this type are of topological origin, the analysis of DoF would be strengthened by explicitly demonstrating that the potential degrees of freedom of the matter sector are entirely gauged away through the gravitational interaction. Thus, it is highly desirable to employ a phase-space approach that systematically provides a robust method to count the number of local physical DoF without resorting to linearization—that is, by taking into account the full set of physical constraints and gauge invariances. This approach would also be helpful in establishing the relationship between the constraint structure and the symmetries of such theories. In this context, a deep understanding of the dynamical content of topological matter models, coupled with $3$d general relativity, requires constructing and analyzing its Hamiltonian formulation \cite{Dirac,Bergmann3}.

The interest of a phase-space treatment here is not merely methodological. Coupling matter to three-dimensional gravity generically introduces local excitations, and with them the very feature that makes the pure theory tractable---the absence of propagating modes---is lost. Models in which the coupling is non-trivial and the topological character nevertheless survives are therefore worth identifying, and whether a given model belongs to that class is not settled by inspecting the action: it is settled by the constraint analysis, and, as we shall see, by a property of the constraint set that a naive count misses. That same analysis is the input of any canonical quantization. In the Dirac programme the first-class constraints are imposed as operator equations annihilating the physical states, and the physical inner product is built on their solution space \cite{Dirac,LQG,Noui}; the second-class ones admit no such treatment and are instead traded for the Dirac bracket, which is what one promotes to a commutator. The split into the two classes thus decides which constraints become operator equations and which become part of the symplectic structure, so that an incomplete classification at the classical level does not stay classical. In the model studied here that second-class sector is not a spectator: it is what deforms the identity obeyed by the first-class constraints, and it must be eliminated before the identity closes. Three-dimensional gravity has long served as the setting in which questions of this kind are settled first, precisely because there they can be settled completely \cite{Ashtekar}.

A Hamiltonian description of any physical system consists of three fundamental components: a phase-space identifying the dynamical degrees of freedom, the Poisson--Dirac brackets defining a symplectic structure, and the Hamiltonian functional, which generates the evolution of the system from given initial conditions by specifying a curve in this phase-space. Nevertheless, as in any covariant theory, the general covariance of the action gives rise to a constrained Hamiltonian system; specifically, the canonical Hamiltonian is not an independent function on phase-space but reduces to a combination of the constraints themselves, a situation familiar from the canonical treatment of topological actions in three and four dimensions \cite{Mignemi}. In this constrained scheme, the dynamics of the system is entirely dictated by the constraints; that is, they define a physical constraint surface within the original phase-space that restricts the system. The subsequent classification of the full set of constraints into first-class and second-class provides the necessary framework to explicitly count the physical DoF. Furthermore, the first-class constraints allow for the construction of the generator of Hamiltonian gauge symmetries—with the prescription of initial data on a Cauchy surface—while the second-class constraints lead to the derivation of the corresponding Dirac brackets. Such a treatment characterizes both the dynamical content and the symmetry structure of the model (see, e.g., Refs. \cite{Henneaux}).

Taking the previously mentioned observations as inspiration, the purpose of this paper is to employ the Dirac--Bergmann protocol to develop in detail a Hamiltonian description of a topological field theory containing a scalar field and an antisymmetric tensor field non-minimally coupled to Einstein--Cartan gravity in three spacetime dimensions \cite{TheModel}. These matter fields are coupled to gravity via a potential that represents a special case of a general class of topological field theories first analyzed by Horowitz \cite{Horowitz}. This model is remarkably rich: it contains unusual black-hole solutions—the so-called exotic BTZ black holes—in which the roles of mass and angular momentum are reversed;  it was shown in Ref. \cite{TheModel} that it takes the form of a generalized Chern--Simons theory once fermionic topological matter is allowed for; and it also admits a description as a higher gauge theory \cite{Girelli,Eugeniu}.  The constrained Hamiltonian analysis à la Dirac--Bergmann presented here provides a complementary view to the minimal phase-space treatment established in Ref. \cite{TheModel}. As has been stressed in the canonical treatment of three-form matter coupled to four-dimensional general relativity \cite{Brizuela}, the advantage of a phase-space description of this kind is that the way the constraints close among themselves already separates the canonical variables into propagating and pure-gauge ones, and that it allows a direct comparison with models carrying other matter contents. Following that same line of reasoning, we hope that this approach will be useful for analyzing the initial-value problem and facilitating the search for numerical solutions. Finally, although beyond the scope of this work, this treatment represents a first step toward the construction of a quantum theory, in which the constraints and their algebra provide the objects to be promoted to operators. Whether the classical symmetry structure survives quantization is a separate question, and one that anomalies may answer in the negative.

Such an analysis is not a purely formal exercise. Coupling matter to gravity may well modify the constraint structure of the vacuum theory: it has been shown, for instance, that a naive matter coupling to a canonically transformed gravitational field turns one of the first-class constraints into a second-class one, and, as a consequence, an unwanted extra mode appears in the phase-space \cite{Aoki}. Part of the interest of the model considered here lies precisely in the fact that this does not occur, and that the reason why it does not can be exhibited explicitly. A closely related situation arises for $p$-form matter in four dimensions, where a three-form field carries four components of which only one turns out to be physical, the remaining ones being fixed by the matter constraints \cite{Brizuela}. On the other hand, the character of a gauge symmetry cannot be established beforehand from the covariant formulation alone. Time reparametrization invariance itself may act trivially on-shell, and hence give rise to no first-class constraint at all, for certain deformations of the Hamiltonian constraint; which of the two situations one is dealing with is decided by the dynamics and by nothing else \cite{Kleinschmidt}. The canonical analysis is what answers the question.

What is new here, and what is not, can be stated precisely. A canonical treatment of this model was initiated by Mann and Popescu \cite{TheModel}. They performed the $(2+1)$ splitting of the action and read the constraints off the temporal components acting as Lagrange multipliers, working directly in the eighteen-dimensional phase-space spanned by the spatial components of the form fields. We shall refer to it as the minimal phase-space. They observed that a naive count of independent constraints returns a negative and therefore inconsistent number of degrees of freedom, identified the redundancy of the gauge parameters responsible for it, and computed the global charges of the theory. Their counting is, by their own statement, carried out in a setting in which no second-class constraints are present; the Legendre transformation and the momenta conjugate to the Lagrange multipliers therefore play no role in it. Completing a canonical treatment along these lines is a well-defined task in its own right. In the study of higher-dimensional Chern--Simons theories the count of local DoF was obtained first. The explicit separation of the second-class constraints, and with it the Dirac bracket that removes them, was carried out afterwards, on the grounds that neither step can be omitted before one attempts to quantize the theory \cite{Banados}. The present work completes the picture in the extended phase-space: we retain the full set of canonical variables, exhibit the eighteen second-class constraints together with the corresponding Dirac brackets, and construct the Hamiltonian gauge generator acting on the entire set of sixty phase-space variables. Along the way, we show that the reducibility of the first-class set follows from an off-shell identity of Ricci type, and that in the extended phase-space it holds only modulo the second-class sector. The mechanism is the following: the constraint that generates the gauge symmetry of the two-form is built from the covariant derivative of the scalar, so that its covariant divergence is a commutator of covariant derivatives, which the Ricci identity returns as the curvature constraint contracted with the scalar itself. That this is a property of the matter content and not of the formalism can be seen by contrast: the same method returns no reducibility conditions for three-dimensional Palatini theory with a cosmological constant \cite{Escalante2014}, nor for the BCEA model of two one-form topological matter fields coupled to gravity \cite{BCEA}.

Our analysis leads to the following picture: coupling topological matter of rank two to Einstein--Cartan gravity renders the algebra of first-class constraints reducible, and the absence of local degrees of freedom characteristic of pure three-dimensional gravity is recovered only once that reducibility is taken into account: treating the twenty-four first-class constraints as independent over-counts them and returns an inconsistent negative value. Throughout this work $\Sigma$ is taken to be a Cauchy surface without boundary. The boundary sector, which requires the Regge--Teitelboim improvement of the constraints \cite{Yekta}, is treated here only in so far as it bears on the reducibility conditions: we show in Sec. \ref{Section5} that the improved generator built from the parameters responsible for them reduces to a single boundary integral, which vanishes precisely when the configuration is topologically trivial. The asymptotic symmetry algebra itself lies outside the scope of this paper and will be addressed separately. The same course has been followed in the Hamiltonian analysis of Palatini Gauss--Bonnet theory, where the conclusion that the model carries neither local nor global degrees of freedom is noted to fail on manifolds with boundaries, and that case is deferred to a separate work \cite{BanadosHenneaux}. We only note here that the dual diffeomorphism charges of three-dimensional gravity, and with them a double Virasoro algebra at any finite distance and without boundary conditions, are revealed only when one works with the connection and the dreibein, and are a priori inaccessible from the Chern--Simons description \cite{Geiller}.

The organization of this article is as follows. In Section \ref{Section2}, we briefly review the Lagrangian formulation of the Einstein--Cartan system with its topological matter content. In Section \ref{Section3}, we perform the $(2+1)$ decomposition of the action to identify the basic variables that span the extended phase-space. The Dirac--Bergmann algorithm is then applied in order to extract every physical constraint of the model and to sort the resulting set according to the brackets it obeys. In Section \ref{Section4}, we construct the gauge-generating functional to derive the complete set of Hamiltonian symmetry transformations and study their relation to spacetime diffeomorphisms and Poincaré transformations. In Section \ref{Section5}, we construct the Dirac brackets that remove the second-class sector, derive the three reducibility conditions obeyed by the first-class set, and use them to count the local degrees of freedom. Finally, in Section \ref{Section6}, we comment on the obtained results and discuss possible future developments.

\section{Lagrangian formulation}
\label{Section2}
We first recall that the Hilbert--Palatini formulation of gravity describes a Lorentz connection soldered to spacetime by the vielbein, on a Riemann--Cartan geometry carrying both curvature and torsion \cite{Blagojevic,Debraj,Utiyama,Kibble}. This should be distinguished from the Chern--Simons formulation of the Poincar\'e group, which describes an unsoldered connection carrying curvature alone. The two are related through the Levi decomposition of the Poincar\'e algebra. All the equations written below belong to the soldered, Hilbert--Palatini side. We note that the term Poincar\'e gauge theory, as used by Kibble, Sciama and in the canonical literature that follows them \cite{Blagojevic,BlagojevicPGT,Utiyama,Kibble}, designates this soldered Riemann--Cartan theory and not the Chern--Simons theory of the Poincar\'e group; it is in that sense that the Poincar\'e transformations recovered in Sec. \ref{Section4} are to be understood. In this formalism, the spacetime metric is no longer a basic gravitational variable but a function of the dreibein $e_{\mu}{^{I}}$ given by $g_{\mu\nu}=\eta_{IJ}e_{\mu}{^{I}}e_{\nu}{^{J}}$, where $\eta_{IJ}$ is the internal Minkowski metric.  The theory carries a local Lorentz invariance acting on the internal index of $e_{\mu}{^{I}}$, and it is the independence of the connection from the vielbein---rather than the presence of local frames, which general relativity also enjoys---that allows torsion to appear and, with it, a coupling to the spin of matter fields. To realize this invariance, covariant derivatives take the place of partial ones, and a spin-connection $A_\mu{^{I}}=1/2\epsilon^{IJK}A_{\mu JK}$, which in three dimensions can be dualized  as shown, is introduced. The dreibein and the connection are now taken as kinematically independent fields, tied to one another by the dynamics alone. The commutator of covariant derivatives yields the curvature $R_{\mu\nu}{^{I}}$, which is the field strength of the connection; the torsion $T_{\mu\nu}{^{I}}$ is not a field strength, since it is not the curvature of any connection, and only on the Chern--Simons side can it be read as the translational component of one. They are defined as
\begin{eqnarray}
    R_{\mu\nu}{^{I}}&=&\partial_{\left[\mu\right.}A_{\left.\nu\right]}{^{I}}+\epsilon^{I}{_{JK}}A_{\mu}{^{J}}A_{\nu}{^{K}},\label{Curvature}\\
    T_{\mu\nu}{^{I}}&=&\mathscr{D}_{[\mu}e_{\nu]}{^{I}},
\end{eqnarray}
where antisymmetrization of the respective indices is denoted by square brackets, e.g., $X_{\left[\mu\right.}Y_{\left.\nu\right]}=X_{\mu}Y_{\nu}-X_{\nu}Y_{\mu}$. Latin letters  $I,J,\dots$ refer to local Lorentz frames, while Greek letters $\mu,\nu,\dots$ refer to coordinate frames. In addition, the covariant derivative of a $SO(2,1)$-valued vector field $X^{I}$ is defined as 
\begin{equation}
    \mathscr{D}_{\mu}X^{I}=\partial_{\mu}X^{I}+\epsilon^{I}{_{JK}}A^{J}_{\mu}X^{K}=\left[\mathscr{D}_{\mu}\right]^{I}{_{K}}X^{K},
\end{equation}
where, for later use, we have defined a differential operator, 
\begin{equation}
    \left[\mathscr{D}_{\mu}\right]^{I}{_{K}}=\delta^{I}_{K}\partial_{\mu}+\epsilon^{I}{_{JK}}A^{J}_{\mu},\label{Operator}
\end{equation}
with $\partial_{\mu}$ being the fiducial derivative operator and $\epsilon^{IJK}$ the completely antisymmetric $SO(2,1)$ tensor. We also denote by $\varepsilon^{\mu\nu\gamma}$ the Levi-Civita tensor density of weight $+1$, normalized so that $\varepsilon^{012}=1$, and by $\varepsilon_{\mu\nu\gamma}$ its counterpart of weight $-1$, fixed by $\varepsilon_{012}=1$. With this normalization the two coincide, in components, with one and the same permutation symbol, and $\varepsilon_{0ab}$ below stands for $\varepsilon_{\mu\nu\gamma}$ with the time index set to zero. The Riemann--Cartan geometry reduces to Riemannian geometry upon the imposition of the zero torsion condition. In this notation, the vacuum dynamics is generated by the action \cite{Carlip}
\begin{equation}
S[e,A]= \int\mathrm{d}^{3}x\varepsilon^{\mu\nu\gamma}\cfrac{1}{2}e_{\mu}{^{I}}R_{\nu\gamma I}.
\label{action1}
\end{equation}
This is the first-order formulation of three dimensional gravity, which is manifestly locally Lorentz invariant in addition to its manifest invariance under diffeomorphisms. Here the dreibein and the connection are taken as independent variables. Varying the action (\ref{action1}) with respect to $e_{\mu}{^{I}}$ and $A_{\mu}{^{I}}$, we obtain the gravitational field equations:
\begin{equation}
\varepsilon^{\mu\nu\gamma}R_{\nu\gamma }{^{I}}=0\quad\text{and}\quad
\varepsilon^{\mu\nu\gamma}T_{\nu\gamma}{^{I}}=0,
\label{Dynamics}
\end{equation}
which state that the spin-connection is locally flat and that, in the absence of matter, the spacetime has vanishing torsion, ensuring that the connection $A_{\mu}{^{I}}$ is compatible with the dreibein $e_{\mu}{^{I}}$, which we will call $A[e]_{\mu}{^{I}}$. Inserting the solution of the second equation into the first one, one obtains Einstein’s equation, $G_{\mu\nu}=0$, where $G_{\mu\nu}$ is the Einstein tensor. This step, and with it the passage to Einsteinian gravity, requires the vielbein to be invertible \cite{Sadovski}.  

As explained in \cite{TheModel}, one can then extend the action (\ref{action1}) to a gravitating topological matter action by introducing a second-rank antisymmetric tensor field $B_{\mu\nu}{^{I}}$, which, in turn, is coupled to  a scalar field $\Phi^{I}$ through the connection $A_{\mu}{^{I}}$ in the covariant derivative. The classical action functional for this topological matter-gravity system is given by
\begin{equation}
S[e,A,B,\Phi]= \int\mathrm{d}^{3}x\varepsilon^{\mu\nu\gamma}\left(\frac{1}{2}e_{\mu}{^{I}}R_{\nu\gamma I}+2B{_{\mu\nu}}^{I}\mathscr{D}_{\gamma}\Phi{_{I}}\right).
\label{action2}
\end{equation}
The scalar-tensor coupling is in the topological form of the diffeomorphism invariant theories considered by Horowitz \cite{Horowitz}. The above action carries two sets of gauge symmetries, Poincaré transformations and spacetime diffeomorphisms. They are not independent of one another: Sec. \ref{Section4} recovers both from one and the same set of first-class constraints, through different field-dependent redefinitions of the gauge parameters. The canonical gauge generator must therefore be built out of all the first-class constraints that a Hamiltonian analysis delivers. 

Before we discuss the Hamiltonian analysis, let us review the equations of motion of this model. Again, the field equations are found by varying the action  (\ref{action2})  with respect to the dynamical fields $e_{\mu}{^{I}}$, $A_{\mu}{^{I}}$, $\Phi^{I}$ and $B_{\mu\nu}{^{I}}$, respectively:
\begin{eqnarray}
\mathcal{E}{^{\mu I}}&:&\cfrac{1}{2}{
\varepsilon^{\mu\nu\gamma}R_{\nu\gamma }{^{I}}}=0,\nonumber\\
\overline{\mathcal{E}}{^{\mu I}}&:&\
\varepsilon^{\mu\nu\gamma}\left(T_{\nu\gamma}{^{I}}-4\epsilon{^{I}}{_{JK}}B_{\nu\gamma}{^{J}}\Phi^{K}\right)=0,\nonumber\\
\mathcal{E}{^{I}}&:&2
\varepsilon^{\mu\nu\gamma}\mathscr{D}_{\mu}B_{\nu\gamma}{^{I}}=0,\nonumber\\
\mathcal{E}^{\mu\nu I}&:&2
\varepsilon^{\mu\nu\gamma}\mathscr{D}_{\gamma}\Phi^{I}=0.\label{EquationMotion}
\end{eqnarray}
It is clear from the first and second equations that the connection $A_{\mu}{^{I}}$ is flat but is generally not compatible with the dreibein $e_{\mu}{^{I}}$. However, this set of equations (\ref{EquationMotion}) determines a non-trivial spacetime geometry; this can be seen by decomposing the full connection $A_{\mu}{^{I}}$ as
\begin{equation}
A_{\mu}{^{I}}=K_{\mu}{^{I}}+\Omega_{\mu}{^{I}},\label{Torsion}
\end{equation}
in terms of a contortion tensor $K_{\mu}{^{I}}$ plus a non-flat spin connection $\Omega_{\mu}{^{I}}$ fixed by the vanishing-torsion condition:
\begin{equation}
\partial_{\left[\mu\right.}e_{\left.\nu\right]}{^{I}}+\epsilon^{I}{_{JK}}\Omega_{\left[\mu\right.}{^{J}} e_{\left.\nu\right]}{^{K}}=0,\label{Freetorsion}    
\end{equation}
which, under the assumption that the dreibein $e_{\mu}{^{I}}$ is invertible with inverse $e_{I}{^{\mu}}$, can be solved for $\Omega_{\mu}{^{I}}$ in terms of $e_{\mu}{^{I}}$, giving the standard relation
\begin{equation}
    \Omega_{\mu}{^{I}}[e]=-e^{-1}\varepsilon^{\nu\rho\sigma}\left(e_{\nu}{^{I}}e_{\mu J}-\cfrac{1}{2}{e_{\mu}{^{I}}e_{\nu J}}\right)\partial_{\rho}e_{\sigma}{^J},\label{NewConnection}
\end{equation}
with $e^{-1}\varepsilon^{\mu\nu\rho}=\epsilon^{IJK}e^{\mu}{_{I}}e^{\nu}{_{J}}e^{\rho}{_{K}}$ and $e=\det e_{\mu}{^{J}}$. Now, the contortion tensor can be solved for by using the second equation in (\ref{EquationMotion}) and Eqs. (\ref{Torsion}) and (\ref{Freetorsion}); one finds that
\begin{equation}
    K_{\mu}{^{I}}[e,B,\Phi]=-4e^{-1}\varepsilon^{\nu\rho\sigma}e_{\rho}{^{I}}\epsilon_{JKL} B_{\mu\nu}{^{J}}e_{\sigma}{^{K}}\Phi^{L}-e_{\mu}{^{I}}e^{-1}\varepsilon^{\nu\rho\sigma}\epsilon_{JKL}B_{\nu\rho}{^{J}}e_{\sigma}{^{K}}\Phi^{L}.\label{ContortionFinal}
\end{equation}
Note that the decomposition (\ref{Torsion}) allows the curvature $R_{\mu\nu}{^{I}}[A]$ to be related to the Riemannian one $R_{\mu\nu}{^{I}}[\Omega]$ of $\Omega[e]_{\mu}{^{I}}$. We have
\begin{eqnarray}
    R_{\mu\nu}{^{I}}[A]&=&R_{\mu\nu}{^{I}}[\Omega]+\mathscr{D}[\Omega]_{[\mu}K_{\nu]}{^{I}}+\epsilon^{I}{_{JK}}K_{\mu}{^{J}}K_{\nu}{^{K}},
\end{eqnarray}
where $\mathscr{D}[\Omega]_{\mu}K_{\nu}{^{I}}=\partial_{\mu}K_{\nu}{^{I}}+\epsilon^{I}{_{JK}}\Omega_{\mu}{^{J}}K_{\nu}{^{K}}$ denotes the covariant derivative with respect to the connection $\Omega_{\mu}{^{I}}$. Under these circumstances, the Riemann tensor receives a shift that can be regarded as the energy-momentum contribution of topological matter to the Einstein equations \cite{Mielke,Vasilic}. Substituting (\ref{NewConnection}) and (\ref{ContortionFinal}) back into (\ref{Torsion}), we finally find
\begin{equation}
    A_{\mu}{^{I}}[e,B,\Phi]=e^{-1}\varepsilon^{\nu\rho\sigma}\left(-4e_{\rho}{^{I}}\epsilon_{JKL} B_{\mu\nu}{^{J}}e_{\sigma}{^{K}}\Phi^{L}-e_{\mu}{^{I}}\epsilon_{JKL}B_{\nu\rho}{^{J}}e_{\sigma}{^{K}}\Phi^{L}-\left(e_{\nu}{^{I}}e_{\mu J}-\cfrac{1}{2}{e_{\mu}{^{I}}e_{\nu J}}\right)\partial_{\rho}e_{\sigma}{^J}\right).
\end{equation}
 A flat Lorentz connection carrying a non-vanishing torsion is also the setting of the torsional deformations of locally AdS$_{3}$ space, where the torsion is decomposed into its irreducible components with respect to the Lorentz structure group and the analysis is restricted to the singlet and the vector, the symmetric-traceless component being left for future study \cite{Zanelli2023}. Equations (\ref{EquationMotion}) produce that configuration on a different footing: the flatness is a field equation rather than a working restriction, the singlet is a function of the fields rather than a constant traded for a cosmological constant, and the torsion, instead of being posited, is fixed by the matter source $\epsilon^{I}{_{JK}}B_{\mu\nu}{^{J}}\Phi^{K}$ and vanishes with it. Wherever that source is non-zero and the dreibein is invertible, the three irreducible components of the torsion are in general all non-vanishing, and the symmetric-traceless one sweeps its entire five-dimensional representation as the two-form varies at a point. The model described by the action (\ref{action2}) can now be recognized as a theory of $3$d gravity. Its variables are a dreibein $e_{\mu}{^{I}}$ and a torsionless but non-flat spin connection $\Omega[e]_{\mu}{^{I}}$, coupled non-trivially to the topological matter fields $\Phi^{I}$ and $B_{\mu\nu}{^{I}}$. The connection $A_{\mu}{^{I}}$ thus encodes two things at once. Its Levi-Civita part carries the metric geometry; its contortion part carries the matter content. 

Some precision is needed about the comparison being made here. The distinction between first- and second-order formalisms refers to the order of the field equations, whereas the distinction between holonomic and non-holonomic variables refers to whether one works with $(g,\Gamma)$ or with $(e,A)$ \cite{Sadovski}. What the non-holonomic variables buy us is that the topological matter of interest couples through the connection alone, without the dreibein, and that geometries with torsion are described directly. This is what makes the canonical treatment worth the effort in the presence of torsion. We now proceed to analyze the constrained Hamiltonian dynamics of (\ref{action2}) using the Dirac--Bergmann procedure \cite{Henneaux,Dirac,Bergmann3}. The analysis will reveal the complete constraint structure of the theory and fix the number of physical degrees of freedom.

%%%%%%%%%%%%%%%%%%%%%%%
\section{Hamiltonian formalism and the Dirac--Bergmann algorithm}
\label{Section3}
Let us now perform the Hamiltonian analysis of the action (\ref{action2}). A word on terminology is in order before we start. Throughout this section, expressions such as ``time evolution'' or ``consistency under evolution'' refer to the flow generated by the Hamiltonian within the Dirac--Bergmann algorithm, which is the standard procedure for extracting the full set of constraints. Since the Hamiltonian of the present model is a combination of constraints, that flow is pure gauge, and no invariant notion of time is being invoked. Systems of this kind are known as totally constrained, or Hamiltonian-constrained \cite{Corichi,Dapor}: the Hamiltonian vector field is annihilated by the symplectic form on the constraint surface, so that its projection to the physical phase-space vanishes identically. Canonical general relativity is the standard example \cite{Dapor}. Introducing a genuine notion of evolution in such a system requires additional structure, provided by the relational, or partial, observables built with a family of clock functions. Moreover, the input of that construction is precisely what a Hamiltonian analysis provides: the constraint surface, the orbits generated by the first-class constraints, and the Dirac bracket induced on the reduced phase-space \cite{Dapor}. The analysis carried out below is, in this sense, not an end in itself. We assume that spacetime has topology $\mathcal{M}=\mathbb{R} \times\Sigma$, where $\Sigma$ denotes a Cauchy surface with $\partial\Sigma=\emptyset$ and $\mathbb{R}$ plays the role of an evolution parameter. We assume, further, that all fields admit simultaneous proper ($2+1$) decompositions. Splitting the indices into their space and time parts, we arrive at the canonical form of the action, $S=\int L\mathrm{d}t$, where $L$ is the Lagrangian function
\begin{equation}
L=\int\mathrm{d}^{2}x \varepsilon^{0ab}\left(\dot{A}_{a}{^{I}}e_{bI}+2B_{ab}{^{I}}\dot{\Phi}_{I}+\cfrac{1}{2}e_{0}{^{I}}R_{abI}+4B_{0a}{^{I}}\mathscr{D}_{b}\Phi_{I}+A_{0}{^{I}}\left(\mathscr{D}_{a}e_{bI}-2\epsilon^{I}{_{JK}}B_{ab}{^{J}}\Phi^{K}\right)\right),\label{Lagrangian}
\end{equation}
where  $a,b,\ldots$ are spatial indices, and the dot stands for the derivative with respect to time. In these expressions the spatial curvature and the covariant derivative read: $R_{ab}{^{I}}=\partial_{[a}A_{b]}{^{I}}+\epsilon^{I}{_{JK}}A_{a}{^{J}}A_{b}{^{K}}$ and $\left[\mathscr{D}_{a}\right]^{I}{_{K}}=\delta^{I}_{K}\partial_{a}+\epsilon^{I}{_{JK}}A^{J}_{a}$. Let us start with a configuration space $\mathcal{Q}$ spanned by all the fields $q^{m}$ that occur in the Lagrangian (\ref{Lagrangian}) and not only by those variables that occur with time derivatives; i.e., $q^{m}[x]=\left(A_{0}{^{I}}, A_{a}{^{I}}, e_{0}{^{I}}, e_{a}{^{I}}, B_{0a}{^{I}}, B_{ab}{^{I}}, \Phi^{I}\right)\in\mathcal{Q}$. Here, and henceforth, the indices $m,n,\dots$ formally denote different kinds of variables, as well as Lorentz and tensorial indices. Then we compute the momenta, denoted as $p_{m}[x]=\left(\Pi^{0}{_{I}},\Pi^{a}{_{I}},\pi^{0}{_{I}},\pi^{a}{_{I}},\Pi^{0a}{_{I}},\Pi^{ab}{_{I}},\Pi_{I}\right)$, canonically conjugate to $q^{m}$ according to 
\begin{eqnarray}
\pi^{0}{_{I}}=\cfrac{\delta{L}}{\delta \dot{e}_{0}{^{I}}}=0,&&\quad\pi^{a}{_{I}}=\cfrac{\delta{L}}{\delta \dot{e}_{a}{^{I}}}=0,\nonumber\\
\Pi^{0}{_{I}}=\cfrac{\delta{L}}{\delta \dot{A}_{0}{^{I}}}=0,&&\quad\Pi^{a}{_{I}}=\cfrac{\delta{L}}{\delta \dot{A}_{a}{^{I}}}=\varepsilon^{0ab}e_{bI},\nonumber\\
\Pi^{0a}{_{I}}=\cfrac{\delta{L}}{\delta \dot{B}_{0a}{^{I}}}=0,&&\quad\Pi^{ab}{_{I}}=\cfrac{\delta{L}}{\delta \dot{B}_{ab}{^{I}}}=0,\nonumber\\
&&\quad\Pi_{I}=\cfrac{\delta{L}}{\delta \dot{\Phi}^{I}}=2\varepsilon^{0ab}B_{abI}.
\label{momentum2}
\end{eqnarray}
The thirty  variables $q^{m}$ as well as their conjugate momenta $p_{m}$ span an extended phase-space $\Gamma$. On $\Gamma$ there exist a Hamiltonian function and a Poisson bracket structure, which together give rise to the dynamics of the model. The Poisson bracket for any pair of functions $\mathcal{F}$ and $\mathcal{G}$ on $\Gamma$ is defined at a fixed time $t$ as 
\begin{equation}
    \{\mathcal{F}[x],\mathcal{G}[y]\}=\int d^{2}z\left(\cfrac{\delta \mathcal{F}[x]}{\delta q^{m}[z]}\cfrac{\delta \mathcal{G}[y]}{\delta p_{m}[z]}-\cfrac{\delta \mathcal{G}[y]}{\delta q^{m}[z]}\cfrac{\delta \mathcal{F}[x]}{\delta p_{m}[z]}\right),\label{DefPoisson-Bracket}
\end{equation}
in such a way that, when applied to the phase-space field variables themselves, equation (\ref{DefPoisson-Bracket}) gives the canonical equal-time Poisson-bracket relations:
\begin{eqnarray}
\{e^{\alpha}{_{I}}[x],\pi_{\beta}{^{J}}[y]\}&=&\delta^{\alpha}_{\beta}\delta^{J}_{I}\delta^{2}(x-y),\nonumber\\
\{A^{\alpha}{_{I}}[x],\Pi_{\beta}{^{J}}[y]\}&=&\delta^{\alpha}_{\beta}\delta^{J}_{I}\delta^{2}(x-y),\nonumber\\
\{B^{\alpha\beta}{_{I}}[x],\Pi_{\mu\nu}{^{J}}[y]\}&=&\cfrac{1}{2}\delta^{\alpha}_{\left[\mu\right.}\delta^{\beta}_{\left.\nu\right]}\delta^{J}_{I}\delta^{2}(x-y),\nonumber\\
\{\Phi_{I}[x],\Pi^{J}[y]\}&=&\delta^{J}_{I}\delta^{2}(x-y),\label{PoissonBracket}
\end{eqnarray}
with $\delta^{2}(x-y)$ being the two-dimensional Dirac delta function.

 Since none of the momenta (\ref{momentum2}) depend on the time derivatives of the  field variables $q^{m}$, the corresponding Hessian matrix, defined as $W_{mn}=\partial p_{m}/\partial\dot{q}^{n}$, is identically zero. Consequently, the rank of $W_{mn}$ is zero; thus, none of the momenta $p_{m}=\delta L/\delta\dot{q}^{m}$ can be solved for the corresponding `velocities' $\dot{q}^{m}$, so that all of them give rise to the following thirty primary constraints:
\begin{eqnarray}
\xi^{0}{_{I}}=\pi^{0}{_{I}}\approx0,&\quad&\xi^{a}{_{I}}=\pi^{a}{_{I}}\approx0,\nonumber\\
\phi^{0}{_{I}}=\Pi^{0}{_{I}}\approx0,&&\phi^{a}{_{I}}=\Pi^{a}{_{I}}-\varepsilon^{0ab}e_{bI}\approx0,\nonumber\\
\psi^{0a}{_{I}}=\Pi^{0a}{_{I}}\approx0,&&\psi^{ab}{_{I}}=\Pi^{ab}{_{I}}\approx0,\nonumber\\
&&\theta_{I}=\Pi_{I}-2\varepsilon^{0ab}B_{abI}\approx0,\label{primary2}
\end{eqnarray}
and they define a constraint surface $\Gamma_{1}\subset\Gamma$. In this manner, the dynamics of our system takes place not on $\Gamma$, but on the subspace $\Gamma_{1}$. Points outside $\Gamma_{1}$ are not accessible. We nevertheless continue to work in $\Gamma$, whose symplectic structure is the one we know; that of the final phase-space is not yet available. The symbol ``$\approx$'' indicates a weak equality in the Dirac sense \cite{Dirac}. That is, we set the constraints equal to zero only when they appear outside of Poisson brackets.

In general, the bracket of any two functionals of the fields and their momenta follows from the fundamental relations (\ref{PoissonBracket}). Accordingly, we calculate the Poisson brackets between the primary constraints (\ref{primary2}), which are as follows:
\begin{eqnarray}
\{\xi^{a}{_{I}}[x],\phi^{b}{_{J}}[y]\}&=&-\varepsilon^{0ab}\eta_{IJ}\delta^{2}(x-y),\nonumber\\
\{\psi^{ab}{_{I}}[x],\theta_{J}[y]\}&=&2\varepsilon^{0ab}\eta_{IJ}\delta^{2}(x-y),\label{PrimaryConstraintAlgebra}
\end{eqnarray}
while all other Poisson brackets vanish. In the expressions that follow, $\left[\mathscr{D}^{\mathbf{x}}_{b}\right]_{IJ}$ denotes the operator (\ref{Operator}) built with the connection evaluated at the point $\mathbf{x}$, acting to the right on the delta function.

Continuing with the Dirac--Bergmann algorithm, consistency demands that the constraint surface $\Gamma_{1}$ be preserved along the evolution of the system. Otherwise further constraints must exist. To explore this, let us construct the so-called primary Hamiltonian, which turns out to be the canonical  Hamiltonian improved by a linear combination of the primary constraints. First, by performing the Legendre transformation of the Lagrangian (\ref{Lagrangian}), we obtain the canonical Hamiltonian:
\begin{equation}
H=-\int\mathrm{d}^{2}x \varepsilon^{0ab}\left[\cfrac{1}{2}e_{0}{^{I}}R_{abI}+4B_{0a}{^{I}}\mathscr{D}_{b}\Phi_{I}+A_{0}{^{I}}\left(\mathscr{D}_{a}e_{bI}-2\epsilon^{I}{_{JK}}B_{ab}{^{J}}\Phi^{K}\right)\right].\label{CHamiltonian2}
\end{equation}
This canonical Hamiltonian does not depend on any momenta, but only on fields and their spatial derivatives. Second, letting $\lambda_{m}=\left(\Lambda_{0}{^{I}}, \Lambda_{a}{^{I}}, \lambda_{0}{^{I}}, \lambda_{a}{^{I}}, \lambda_{0a}{^{I}}, \lambda_{ab}{^{I}}, \lambda{^{I}}\right)$ and $\Upsilon^{m}=(\xi^{0}{_{I}},\xi^{a}{_{I}},\phi^{0}{_{I}},\phi^{a}{_{I}},\psi^{0a}{_{I}},\psi^{ab}{_{I}},\theta_{I})$ be the set of Lagrange multipliers and the primary constraints, respectively, we construct the primary Hamiltonian:
\begin{equation}
H_{1}=H+\int \mathrm{d}^{2}x\lambda_{m}[x]\Upsilon^{m}[x].\label{HamiltonTotal}
\end{equation}
In terms of the Hamiltonian (\ref{HamiltonTotal}) and the brackets (\ref{PoissonBracket}), the development of the primary constraints reads 
\begin{equation}
\cfrac{\mathrm{d} \Upsilon^{n}}{\mathrm{d}t}=\{\Upsilon^{n},H_{1}\}\approx\{\Upsilon^{n},H\}+\int \mathrm{d}^{2}x\{\Upsilon^{n},\Upsilon^{m}\}\lambda_{m}[x],    \end{equation}
where we dropped the term $\int \mathrm{d}^{2}x\{\Upsilon^{n},\lambda_{m}\}\Upsilon^{m}[x]$ because it represents a linear combination of constraints. Then we can test the time evolution of the primary constraints according to the primary Hamiltonian $H_{1}$. Consistency requires $\mathrm{d}{\Upsilon}^{m}/\mathrm{d}t\approx0$. This may give new constraints, fix some Lagrange multipliers, or hold identically. We find that the consistency of the primary constraints (\ref{primary2}) leads to the following expressions:
\begin{eqnarray}
\{\xi^{0}{_{I}},H_{1}\}&=&\cfrac{1}{2}\varepsilon^{0ab}R_{abI}\approx0,\nonumber\\
\{\xi^{a}{_{I}},H_{1}\}&=&\varepsilon^{0ab}\left(\mathscr{D}_{b}A_{0I}-\lambda_{bI}\right)\approx0,\nonumber\\
\{\phi^{0}{_{I}},H_{1}\}&=&\varepsilon^{0ab}\left(\mathscr{D}_{a}e_{bI}-2\epsilon_{IJK}B_{ab}{^{J}}\Phi^{K}\right)\approx0,\nonumber\\
\{\phi^{a}{_{I}},H_{1}\}&=&\varepsilon^{0ab}\left(\mathscr{D}_{b}e_{0I}+4\epsilon_{IJK}B_{0b}{^{J}}\Phi^{K}-\epsilon_{IJK}A_{0}{^{J}}e_{bK}-\Lambda_{bI}\right)\approx0,\nonumber\\
\{\psi^{0a}{_{I}},H_{1}\}&=&2\varepsilon^{0ab}\mathscr{D}_{b}\Phi_{I}\approx0,\nonumber\\
\{\psi^{ab}{_{I}},H_{1}\}&=&2\varepsilon^{0ab}\left(\epsilon_{IJK}A_{0}{^{J}}\Phi^{K}+\lambda_{I}\right)\approx0,\nonumber\\
\{\theta_{I},H_{1}\}&=&2\varepsilon^{0ab}\left(2\mathscr{D}_{a}B_{0bI}-\epsilon_{IJK}A_{0}{^{J}}B_{ab}{^{K}}-\lambda_{abI}\right)\approx0.\label{Stability}
\end{eqnarray}
The consistency of $\xi^{0}{_{I}}$, $\phi^{0}{_{I}}$ and $\psi^{0a}{_{I}}$ leads to twelve secondary constraints, namely,
\begin{eqnarray}
\mathfrak{A}_{I}&=&\frac{1}{2}\varepsilon^{0ab}R_{abI}\approx0,\nonumber\\
\mathfrak{B}_{I}&=&\varepsilon^{0ab}\left(\mathscr{D}_{a}e_{bI}-2\epsilon_{IJK}B_{ab}{^{J}}\Phi^{K}\right)\approx0,\nonumber\\
\mathfrak{C}^{a}{_{I}}&=&2\varepsilon^{0ab}\mathscr{D}_{b}\Phi_{I}\approx0.\label{SecondaryConstraints}
\end{eqnarray}
The evolution of $\xi^{a}{_{I}}$, $\phi^{a}{_{I}}$, $\psi^{ab}{_{I}}$ and $\theta_{I}$, in contrast, determines some of the unknown multipliers:
\begin{eqnarray}
\lambda_{bI}&\approx&\mathscr{D}_{b}A_{0I},\nonumber\\
\Lambda_{bI}&\approx&\mathscr{D}_{b}e_{0I}+4\epsilon_{I}{^{JK}}B_{0bJ}\Phi_{K}-\epsilon_{I}{^{JK}}A_{0J}e_{bK},\nonumber\\
\lambda_{I}&\approx&-\epsilon_{I}{^{JK}}A_{0J}\Phi_{K},\nonumber\\
\lambda_{abI}&\approx&2\mathscr{D}_{a}B_{0bI}-\epsilon_{I}{^{JK}}A_{0J}B_{abK}.\label{LagrangeMultipliers}
\end{eqnarray}

In summary, up to now, we have found forty-two constraints:
\begin{eqnarray}
    30\quad\text{Primary constraints}&:&\Upsilon^{m}=(\xi^{0}{_{I}},\xi^{a}{_{I}},\phi^{0}{_{I}},\phi^{a}{_{I}},\psi^{0a}{_{I}},\psi^{ab}{_{I}},\theta_{I})\approx0,\nonumber\\
    12\quad\text{Secondary constraints}&:&\Omega^{m}=(\mathfrak{A}_{I},\mathfrak{B}_{I},\mathfrak{C}^{a}{_{I}})\approx0\label{ObtainedConstraints}.
\end{eqnarray}
Remark that the primary and secondary constraints define a new subspace  $\Gamma_{2}\subseteq\Gamma_{1}$. The dynamics is now confined to the secondary constraint surface $\Gamma_{2}$. Weak equalities are, from here on, understood with respect to $\Gamma_{2}$.  For consistency of the model, we must further ensure that $\Gamma_{2}$ does not change with time. 

Before checking the consistency of the constraints, let us analyze their Poisson-bracket algebra.  First of all, we note that none of the secondary constraints $\Omega^{m}$ (\ref{SecondaryConstraints}) depend on the momenta, so they trivially commute with themselves, but not, in general, with the primary constraints $\xi^{a}{_{I}},\phi^{a}{_{I}},\psi^{ab}{_{I}}, \theta_{I}$, giving us explicitly,
\begin{eqnarray}
\{\mathfrak{C}^{a}{_{I}}[x],\phi^{b}{_{J}}[y]\}&=&2\varepsilon^{0ab}\epsilon_{IJK}\Phi^{K}\delta^{2}(x-y),\nonumber\\
\{\mathfrak{C}^{a}{_{I}}[x],\theta_{J}[y]\}&=&2\varepsilon^{0ab}\left[\mathscr{D}^{\mathbf{x}}_{b}\right]{_{IJ}}\delta^{2}(x-y),\nonumber\\
\{\mathfrak{B}_{I}[x],\xi^{a}{_{J}}[y]\}&=&-\varepsilon^{0ab}\left[\mathscr{D}^{\mathbf{x}}_{b}\right]{_{IJ}}\delta^{2}(x-y),\nonumber\\
\{\mathfrak{B}_{I}[x],\phi^{a}{_{J}}[y]\}&=&\varepsilon^{0ab}\epsilon{_{IJK}}e_{b}{^{K}}\delta^{2}(x-y),\nonumber\\
\{\mathfrak{B}_{I}[x],\psi^{ab}{_{J}}[y]\}&=&-2\varepsilon^{0ab}\epsilon{_{IJK}}\Phi^{K}\delta^{2}(x-y),\nonumber\\
\{\mathfrak{B}_{I}[x],\theta_{J}[y]\}&=&2\varepsilon^{0ab}\epsilon{_{IJK}}B_{ab}{^{K}}\delta^{2}(x-y),\nonumber\\
\{\mathfrak{A}_{I}[x],\phi^{a}{_{J}}[y]\}&=&-\varepsilon^{0ab}\left[\mathscr{D}^{\mathbf{x}}_{b}\right]{_{IJ}}\delta^{2}(x-y).\label{algebra}
\end{eqnarray}
Notice now that, on account of the expressions (\ref{algebra}), none of the secondary constraints, taken as they stand, has vanishing brackets with the whole constraint set, so that none of them is first class on its own. The primary constraints $\xi^{0}{_{I}},\phi^{0}{_{I}},\psi^{0a}{_{I}}$, by contrast, strongly commute with every constraint, including themselves, and are therefore the only manifestly first-class ones at this stage.  Nevertheless, there is the possibility of having some linear combinations of constraints that are first class without affecting the dynamics on the constraint surface $\Gamma_{2}$ \cite{Henneaux,Govaerts}. To investigate this, we concentrate on the matrix $\boxplus_{(xy)}^{mn}$ whose entries are the Poisson brackets involving the primary constraints $\Xi^{m}=(\xi^{a}{_{I}},\phi^{a}{_{I}},\psi^{ab}{_{I}},\theta_{I})$ and all the secondary constraints $\Omega^{m}$, which have already been calculated in (\ref{PrimaryConstraintAlgebra}) and (\ref{algebra}). This matrix has the following form
\begin{equation}
 \boxplus_{(xy)}^{mn}=\begin{pmatrix}
\{\Xi^{m}[x],\Xi^{n}[y]\} & \{\Xi^{m}[x],\Omega^{n}[y]\} \\
\{\Omega^{m}[x],\Xi^{n}[y]\} & \{\Omega^{m}[x],\Omega^{n}[y]\}
\end{pmatrix},\label{MatrixAllConstraints}   
\end{equation}
where the elements corresponding to the Poisson brackets $\{\Omega^{m}[x],\Omega^{n}[y]\}$ are identically  zero. In this case, the above matrix is singular, and therefore not invertible on the constraint surface $\Gamma_{2}$, as the rank count given below shows. This is the geometric content of Dirac's classification: the split into first- and second-class constraints is read off from the degeneracy of the symplectic two-form restricted to the constraint surface, and the gauge-invariant information is what survives the quotient along the orbits of the first-class ones \cite{Corichi}. From the Hamiltonian point of view, the singularity of the matrix (\ref{MatrixAllConstraints}) means that not all constraints $\Delta^{m}=(\Xi^{m},\Omega^{n})$ are second-class, and subsequently, some combinations of these constraints  must be first class  \cite{Govaerts}. To separate explicitly first-class constraints from the rest of the constraints in $\Delta^{m}$, we analyze the properties of the matrix (\ref{MatrixAllConstraints}). Since the matrix (\ref{MatrixAllConstraints}) is singular, it necessarily possesses some independent zero-modes $V_{m}$, which are a solution of the corresponding zero-mode equation $\int\mathrm{d}^{2}x\boxplus_{(xy)}^{mn} V_{n}=0$ on $\Gamma_{2}$. They read,
\begin{eqnarray}
V^{1}_{m}&=&\left(\left[\mathscr{D}^{\mathbf{x}}_{a}\right]_{IJ},0,0,0,\eta_{IJ},0,0\right),\nonumber\\
V^{2}_{m}&=&\left(-\epsilon_{IJK}e_{a}{^{K}},\left[\mathscr{D}^{\mathbf{x}}_{a}\right]_{IJ},-\epsilon_{IJK}B_{ab}{^{K}},-\epsilon_{IJK}\Phi^{K},0,\eta_{IJ},0\right),\nonumber\\
V^{3}_{m}&=&\left(2\epsilon_{IJK}\Phi^{K},0,-\left[\mathscr{D}^{\mathbf{x}}_{b}\right]_{IJ},0,0,0,\eta_{IJ}\right).\label{ZeroModes}
\end{eqnarray}
These twelve families exhaust the kernel, as the following argument shows. By (\ref{PrimaryConstraintAlgebra}), the block $\{\Xi^{m}[x],\Xi^{n}[y]\}$ of (\ref{MatrixAllConstraints}) carries no derivatives of the delta function and is block diagonal in the pairs $(\xi^{a}{_{I}},\phi^{b}{_{J}})$ and $(\psi^{ab}{_{I}},\theta_{J})$. Each block is built out of $\varepsilon^{0ab}$ and $\eta_{IJ}$ alone and is therefore invertible; this $18\times18$ submatrix has maximal rank. Since the rank of a matrix is never smaller than that of any of its submatrices, $\mathrm{rank}\,\boxplus_{(xy)}^{mn}\geq18$, and the kernel has dimension at most $30-18=12$. The modes (\ref{ZeroModes}), on the other hand, are twelve and manifestly independent: their components along $\Omega^{m}$ reduce to $\eta_{IJ}$ in one slot and vanish in the remaining ones. The kernel is therefore exactly twelve dimensional and $\mathrm{rank}\,\boxplus_{(xy)}^{mn}=18$. No further first-class combination can be formed out of $\Delta^{m}$, and the eighteen remaining constraints are second class.

As we shall see, it follows that associated with all zero-modes (\ref{ZeroModes}), the combinations of constraints $\int\mathrm{d}^{2}x\,V^{1,2,3}_{m}\Delta^{m}$ are actually first-class constraints (see, e.g., \cite{Govaerts} for details). These are given explicitly by
\begin{eqnarray}
\widetilde{\mathfrak{A}}_{I}&=&\int\mathrm{d}^{2}x V^{1}_{m}(x)\Delta^{m}(x)=\mathfrak{A}_{I}+\mathscr{D}_{a}\xi^{a}{_{I}},\nonumber\\
\widetilde{\mathfrak{B}}_{I}&=&\int\mathrm{d}^{2}xV_{m}^{2}(x)\Delta^{m}(x)=\mathfrak{B}_{I}+\mathscr{D}_{a}\phi^{a}{_{I}}+\epsilon_{IJK}\left(e_{a}{^{J}}\xi^{a}{^{K}}+\Phi^{J}\theta^{K}+B_{ab}{^{J}}\psi^{ab}{^{K}}\right),\nonumber\\
\widetilde{\mathfrak{C}}_{I}^{a}&=&\int\mathrm{d}^{2}x V_{m}^{3}(x)\Delta^{m}(x)=\mathfrak{C}^{a}{_{I}}-\mathscr{D}_{b}\psi^{ab}{_{I}}-2\epsilon_{IJK}\Phi^{J}\xi^{a}{^{K}}. \label{ModifiedSecondaryConstraints}
\end{eqnarray}
An explicit calculation shows that the modified constraints (\ref{ModifiedSecondaryConstraints}) close among themselves,
\begin{eqnarray}
    \{\widetilde{\mathfrak{A}}_{I}[x],\widetilde{\mathfrak{B}}_{J}[y]\}&=&\epsilon_{IJ}{^{K}}\widetilde{\mathfrak{A}}_{K}\delta^{2}(x-y),\nonumber\\
\{\widetilde{\mathfrak{B}}_{I}[x],\widetilde{\mathfrak{B}}_{J}[y]\}&=&\epsilon_{IJ}{^{K}}\widetilde{\mathfrak{B}}_{K}\delta^{2}(x-y),\nonumber\\
\{\widetilde{\mathfrak{B}}_{I}[x],\widetilde{\mathfrak{C}}^{a}{_{J}}[y]\}&=&\epsilon_{IJ}{^{K}}\widetilde{\mathfrak{C}}^{a}{_{K}}\delta^{2}(x-y).\label{FinalPoissonAlgebra1}
\end{eqnarray}
These constraints must vanish consistently in all gauges, as shown below. This algebra describes the gauge content. It closes with field-independent structure constants, and not with the structure functions of the hypersurface-deformation algebra of general relativity; the diffeomorphisms are recovered only through the field-dependent redefinition (\ref{mapping1}) of the gauge parameters. On the other hand, (\ref{FinalPoissonAlgebra1}) can be compared with the algebra obtained long ago for the version of this class of models in which the scalar field is a Lorentz scalar and the two-form is not valued in the algebra \cite{Gegenberg1}. There the three sets of constraints obey brackets of the same structure as those found here, with a single exception: the generator of the abelian two-form symmetry is there central---it commutes with every constraint, and in particular with the Lorentz generator---whereas here $\widetilde{\mathfrak{C}}^{a}{_{I}}$ transforms in the adjoint representation and one finds $\{\widetilde{\mathfrak{B}}_{I},\widetilde{\mathfrak{C}}^{a}{_{J}}\}=\epsilon_{IJ}{^{K}}\widetilde{\mathfrak{C}}^{a}{_{K}}$. This is what one expects once the matter fields are taken to be $SO(2,1)$-valued. Here $\widetilde{\mathfrak{B}}_{I}$ is the generator of local Lorentz rotations. Every object carrying a single internal index is rotated by it with the same weight, as the brackets (\ref{FinalPoissonAlgebra}) with the second-class constraints also show. Moreover, the non-zero Poisson brackets between them and the primary constraints (\ref{primary2}) take the following form:
\begin{eqnarray}
\{\widetilde{\mathfrak{A}}_{I}[x],\phi^{a}{_{J}}[y]\}&=&\epsilon_{IJ}{^{K}}\xi^{a}{_{K}}\delta^{2}(x-y),\nonumber\\
\{\widetilde{\mathfrak{B}}_{I}[x],\xi^{a}{_{J}}[y]\}&=&\epsilon_{IJ}{^{K}}\xi^{a}{_{K}}\delta^{2}(x-y),\nonumber\\
\{\widetilde{\mathfrak{B}}_{I}[x],\phi^{a}{_{J}}[y]\}&=&\epsilon_{IJ}{^{K}}\phi^{a}{_{K}}\delta^{2}(x-y),\nonumber\\
\{\widetilde{\mathfrak{B}}_{I}[x],\psi^{ab}{_{J}}[y]\}&=&\epsilon_{IJ}{^{K}}\psi^{ab}{_{K}}\delta^{2}(x-y),\nonumber\\
\{\widetilde{\mathfrak{B}}_{I}[x],\theta_{J}[y]\}&=&\epsilon_{IJ}{^{K}}\theta_{K}\delta^{2}(x-y),\nonumber\\
\{\widetilde{\mathfrak{C}}^{a}{_{I}}[x],\phi^{b}{_{J}}[y]\}&=&-\epsilon_{IJ}{^{K}}\psi^{ab}{_{K}}\delta^{2}(x-y),\nonumber\\
\{\widetilde{\mathfrak{C}}^{a}{_{I}}[x],\theta_{J}[y]\}&=&-2\epsilon_{IJ}{^{K}}\xi^{a}{_{K}}\delta^{2}(x-y).\label{FinalPoissonAlgebra}
\end{eqnarray}
For consistency, we must also ensure that the secondary constraint surface $\Gamma_{2}$ is preserved in time evolution: the primary and modified secondary constraints must be constants  of motion on $\Gamma_{2}$ with respect to the so-called secondary Hamiltonian $H_{2}$. Such a secondary Hamiltonian is simply given by the primary Hamiltonian plus a linear combination of all the modified secondary constraints (\ref{ModifiedSecondaryConstraints}), so that the secondary Hamiltonian of our system is just 
\begin{equation}
    H_{2}=H_{1}+\int\mathrm{d}^{2}x\left[\mathfrak{a}_{0}{^{I}}\widetilde{\mathfrak{A}}_{I}+\mathfrak{b}_{0}{^{I}}\widetilde{\mathfrak{B}}_{I}+\mathfrak{c}_{0a}{^{I}}\widetilde{\mathfrak{C}}^{a}{_{I}}\right],
\end{equation}
where $\mathfrak{a}_{0}{^{I}}$, $\mathfrak{b}_{0}{^{I}}$, and $ \mathfrak{c}_{0a}{^{I}}$ are new multipliers for the constraints $\widetilde{\mathfrak{A}}_{I}$, $\widetilde{\mathfrak{B}}_{I}$, and $\widetilde{\mathfrak{C}}^{a}{_{I}}$, respectively. Nevertheless, using Eqs. (\ref{LagrangeMultipliers}), the above Hamiltonian takes the form
\begin{equation}
H_{2}=\int\mathrm{d}^{2}x\left[\left(\mathfrak{a}_{0}{^{I}}-e_{0}{^{I}}\right)\widetilde{\mathfrak{A}}_{I}+ \left(\mathfrak{b}_{0}{^{I}}-A_{0}{^{I}}\right)\widetilde{\mathfrak{B}}_{I}+\left(\mathfrak{c}_{0a}{^{I}}-2B_{0a}{^{I}}\right)\widetilde{\mathfrak{C}}^{a}{_{I}}+\Lambda_{0}{^{I}}\xi^{0}{_{I}}+\lambda_{0}{^{I}}\phi^{0}{_{I}}+\lambda_{0a}{^{I}}\psi^{0a}{_{I}}\right].\end{equation}
It follows that the fields $e_{0}{^{I}}$, $A_{0}{^{I}}$, and $B_{0a}{^{I}}$ also represent a set of multipliers for the modified constraints $\widetilde{\mathfrak{A}}_{I}$, $\widetilde{\mathfrak{B}}_{I}$, and $\widetilde{\mathfrak{C}}^{a}{_{I}}$, so that no further multipliers need be introduced. Setting $\mathfrak{a}_{0}{^{I}}=\mathfrak{b}_{0}{^{I}}=\mathfrak{c}_{0a}{^{I}}=0$ therefore returns the canonical Hamiltonian (\ref{CHamiltonian2}) written in terms of the modified constraints, so that the secondary Hamiltonian boils down to
\begin{equation}
H_{2}=\int\mathrm{d}^{2}x\left[-e_{0}{^{I}}\widetilde{\mathfrak{A}}_{I}-A_{0}{^{I}}\widetilde{\mathfrak{B}}_{I}-2B_{0a}{^{I}}\widetilde{\mathfrak{C}}^{a}{_{I}}+\Lambda_{0}{^{I}}\xi^{0}{_{I}}+\lambda_{0}{^{I}}\phi^{0}{_{I}}+\lambda_{0a}{^{I}}\psi^{0a}{_{I}}\right].\label{FinalHamiltonian}
\end{equation}
Armed with these facts, the preservation of all constraints over time can be checked by taking their Poisson brackets with $H_{2}$. Specifically, we show that,
\begin{eqnarray}
    \{\xi^{0}{_{I}},H_{2}\}&=&\widetilde{\mathfrak{A}}_{I}\approx0,\nonumber\\
    \{\xi^{a}{_{I}},H_{2}\}&=&-\epsilon_{IJ}{^{K}}A_{0}{^{J}}\xi^{a}{_{K}}\approx0,\nonumber\\
    \{\phi^{0}{_{I}},H_{2}\}&=&\widetilde{\mathfrak{B}}_{I}\approx0,\nonumber\\
    \{\phi^{a}{_{I}},H_{2}\}&=&-\epsilon_{IJ}{^{K}}\left(e_{0}{^{J}}\xi^{a}{_{K}}+2B_{0b}{^{J}}\psi^{ab}{_{K}}+A_{0}{^{J}}\phi^{a}{_{K}}\right)\approx0,\nonumber\\
    \{\psi^{0a}{_{I}},H_{2}\}&=&\widetilde{\mathfrak{C}}^{a}{_{I}}\approx0,\nonumber\\
    \{\psi^{ab}{_{I}},H_{2}\}&=&-\epsilon_{IJ}{^{K}}A_{0}{^{J}}\psi^{ab}{_{K}}\approx0,\nonumber\\
    \{\theta_{I},H_{2}\}&=&-\epsilon_{IJ}{^{K}}\left(A_{0}{^{J}}\theta_{K}-4B_{0a}{^{J}}\xi^{a}{_{K}}\right)\approx0,\nonumber\\
\{
\widetilde{\mathfrak{A}}_{I},H_{2}\}&=&-\epsilon_{IJ}{^{K}}A_{0}{^{J}}\widetilde{\mathfrak{A}}_{K}\approx0,\nonumber\\
    \{\widetilde{\mathfrak{B}}_{I},H_{2}\}&=&-\epsilon_{IJ}{^{K}}\left(A_{0}{^{J}}\widetilde{\mathfrak{B}}_{K}+2B_{0a}{^{J}}\widetilde{\mathfrak{C}}^{a}{_{K}}+e_{0}{^{J}}\widetilde{\mathfrak{A}}_{K}\right)\approx0,\nonumber\\
    \{\widetilde{\mathfrak{C}}^{a}{_{I}},H_{2}\}&=&-\epsilon_{IJ}{^{K}}A_{0}{^{J}}\widetilde{\mathfrak{C}}^{a}{_{K}}\approx0.
\end{eqnarray}
The consistency conditions generated by $H_{2}$ are satisfied without further requirements. There are no tertiary constraints. Consequently, the Dirac--Bergmann algorithm stops.

Having determined the structure of all physical constraints in the theory, we now classify them into first- and second-class. First-class constraints commute, at least weakly, with every constraint in the system. They are generally associated with an internal gauge symmetry acting at each spacetime point. Further, according to Dirac's prescription for the quantization of constrained systems, these first-class constraints are implemented as operators that annihilate the wave function \cite{Corichi}. Those with at least one weakly non-vanishing bracket are second class. They remove redundant phase-space variables.

It is easy to see from the constraint algebra given above in (\ref{PrimaryConstraintAlgebra}), (\ref{FinalPoissonAlgebra1}), and (\ref{FinalPoissonAlgebra}), that all Poisson brackets of constraints $\widetilde{\mathfrak{A}}_{I},\,
    \widetilde{\mathfrak{B}}_{I},\, \widetilde{\mathfrak{C}}^{a}{_{I}},\, \xi^{0}{_{I}},\, \phi^{0}{_{I}},\, \psi^{0a}{_{I}}$ vanish on $\Gamma_{2}$, so that the following set is first class:
\begin{equation}
    \text{$24$ first-class}\,:\,\gamma^{m}=\left( \widetilde{\mathfrak{A}}_{I},\, \widetilde{\mathfrak{B}}_{I},\,
    \widetilde{\mathfrak{C}}^{a}{_{I}},\, \xi^{0}{_{I}},\, \phi^{0}{_{I}},\, \psi^{0a}{_{I}}\right)\approx0. \label{FirstClass}
\end{equation}
This means that the remaining constraints are second-class ones:
\begin{equation}
    \text{$18$ second-class}\,:\,\chi^{m}=\left(\xi^{a}{_{I}},\,\phi^{a}{_{I}},\,\psi^{ab}{_{I}},\,\theta_{I}\right)\approx0.\label{SecondClass}
\end{equation}
This concludes the constraint analysis. The sections that follow put both classes to work: the first class in identifying gauge symmetries, the second in counting the physical degrees of freedom.
%%%%%%%%%%%%%%%%%%%%%%%%%%%%%%%%%%%%%%%%%%%%%%%%%%%%%%
\section{First-class constraints and gauge transformations}
\label{Section4}
We are now ready to calculate the generator of the gauge symmetries of our model. As we know, gauge invariance plays a central role in our understanding of the fundamental interactions. Gauge symmetry is also what makes their renormalizable description possible \cite{Hooft}, and it restricts what may count as an observable, since a physical quantity must be invariant under the gauge transformations \cite{Dirac,Henneaux}. According to the Dirac prescription, the Hamiltonian gauge symmetries parametrized at the infinitesimal level by $\alpha_{m}$ are generated by a linear combination of all first-class constraints $\gamma^{m}$ \cite{Dirac}. Thus, the expression for the generating functional of gauge transformations can be written as 
\begin{equation}
    G=\int\mathrm{d}^{2}x \alpha_{m}[x]\gamma^{m}[x].
\end{equation}
However, not all the gauge parameters $\alpha_{m}$ are independent. The number of independent gauge parameters is equal to the number of independent primary first-class constraints; in a reducible theory this is the count prior to the reducibility conditions, a point we return to in Sec. \ref{Section5}. One can systematically eliminate the dependent ones and construct the Hamiltonian generator of the gauge symmetries in terms of only the independent $\alpha_{m}$, following any of the procedures given in \cite{Banerjee,Pons, Castellani}.  We employ the Castellani procedure \cite{Henneaux,Castellani} to construct the Hamiltonian gauge generator. Starting with the primary first-class constraints $\xi^{0}{_{I}}$, $\phi^{0}{_{I}}$, and $\psi^{0a}{_{I}}$, we find:
\begin{eqnarray}
    \mathcal{G}[\sigma]&=&-\dot{\sigma}^{I}\xi^{0}{_{I}}+\sigma^{I}\left(\widetilde{\mathfrak{A}}_{I}+\epsilon_{IJ}{^{K}}A_{0}{^{J}}\xi^{0}{_{K}}\right),\nonumber\\
\mathcal{G}[\omega]&=&-\dot{\omega}^{I}\phi^{0}{_{I}}+\omega^{I}\left(\widetilde{\mathfrak{B}}_{I}+\epsilon_{IJ}{^{K}}\left(A_{0}{^{J}}\phi^{0}{_{K}}+e_{0}{^{J}}\xi^{0}{_{K}}+2B_{0a}{^{J}}\psi^{0a}{_{K}}\right)\right),\nonumber\\
\mathcal{G}[\rho]&=&-\dot{\rho}_{a}{^{I}}\psi^{0a}{_{I}}+\rho_{a}{^{I}}\left(\widetilde{\mathfrak{C}}^{a}{_{I}}
+\epsilon_{IJ}{^{K}}A_{0}{^{J}}\psi^{0a}{_{K}}\right),
\end{eqnarray}
where $\sigma^{I}$, $\omega^{I}$, and $\rho_{a}{^{I}}$ are the time-dependent gauge parameters. The complete Hamiltonian generator of the gauge transformation has the form
\begin{equation}
G=\int \mathrm{d}^{2}y\left(\mathcal{G}[\sigma]+\mathcal{G}[\omega]+\mathcal{G}[\rho]\right).\label{ParticularGenerator}
\end{equation}
Hence, its action on the variables $q^{m}$ and their conjugate momenta $p_{m}$ is defined by the Poisson bracket\footnote{ Or a Dirac bracket, once the second-class constraints have been eliminated \cite{Henneaux}.} operation with the corresponding fields,
\begin{equation}
    \delta_{G}q^{m}=\{q^{m},G\}\quad\text{and}\quad\delta_{G}p_{m}=\{p_{m},G\}.
\end{equation}
After some calculations, we arrive at the symmetry transformations on the full set of canonical variables. Those of the fields read
\begin{eqnarray}
\delta_{G} e_{0}{^{I}}&=&-\mathscr{D}_{0}\sigma^{I}+\epsilon^{I}{_{JK}}\omega^{J}e_{0}{^{K}},\nonumber\\
\delta_{G} e_{a}{^{I}}&=&-\mathscr{D}_{a}\sigma^{I}+\epsilon^{I}{_{JK}}\left(\omega^{J}e_{a}{^{K}}-2\rho_{a}{^{J}}\Phi^{K}\right),\nonumber\\
\delta_{G} A_{0}{^{I}}&=&-\mathscr{D}_{0}\omega^{I},\nonumber\\
\delta_{G} A_{a}{^{I}}&=&-\mathscr{D}_{a}\omega^{I},\nonumber\\
\delta_{G} B_{ab}{^{I}}&=&-\cfrac{1}{2}\mathscr{D}_{\left[a\right.}\rho_{\left.b\right]}{^{I}}-\epsilon^{I}{_{JK}}B_{ab}{^{J}}\omega^{K},\nonumber\\
\delta_{G} B_{0a}{^{I}}&=&-\cfrac{1}{2}\mathscr{D}_{0}\rho_{a}{^{I}}-\epsilon^{I}{_{JK}}B_{0a}{^{J}}\omega^{K},\nonumber\\
\delta_{G} \Phi^{I}&=&\epsilon^{I}{_{JK}}\omega^{J}\Phi^{K}.\label{HamltonianGaugeSymmetries}
\end{eqnarray}
The variations of the momenta follow from the same bracket:
\begin{eqnarray}
\delta_{G} \pi_{0}{^{I}}&=&\epsilon^{I}{_{JK}}\omega^{J}\pi_{0}{^{K}},\nonumber\\
\delta_{G} \pi_{a}{^{I}}&=&\epsilon^{I}{_{JK}}\omega^{J}\pi_{a}{^{K}},\nonumber\\
\delta_{G} \Pi_{0}{^{I}}&=&\epsilon^{I}{_{JK}}\left(\sigma^{J}\pi_{0}{^{K}}+\omega^{J}\Pi_{0}{^{K}}+\rho_{b}{^{J}}\psi_{0}{^{bK}}\right),\nonumber\\
\delta_{G} \Pi_{a}{^{I}}&=&-\varepsilon_{0ab}\left(\mathscr{D}^{b}\sigma^{I}+2\epsilon^{I}{_{JK}}\rho^{bJ}\Phi^{K}\right)+\epsilon^{I}{_{JK}}\left(\sigma^{J}\pi_{a}{^{K}}+\omega^{J}\Pi_{a}{^{K}}+\rho_{b}{^{J}}\psi_{a}{^{bK}}\right),\nonumber\\
\delta_{G} \Pi_{ab}{^{I}}&=&-\epsilon^{I}{_{JK}}\Pi_{ab}{^{J}}\omega^{K},\nonumber\\
\delta_{G} \Pi_{0a}{^{I}}&=&-\epsilon^{I}{_{JK}}\Pi_{0a}{^{J}}\omega^{K},\nonumber\\
\delta_{G} \Pi^{I}&=&-2\varepsilon^{0ab}\mathscr{D}_{a}\rho_{b}{^{I}}+\epsilon^{I}{_{JK}}\left(\omega^{J}\Pi^{K}-2\rho_{a}{^{J}}\pi^{aK}\right).\label{MomentaGaugeSymmetries}
\end{eqnarray}
Since the whole set of canonical variables is retained, these infinitesimal transformations are new; they are realized on the full set of phase-space variables. Two remarks are in order before we proceed. First, the independent gauge parameters are as many as the independent primary first-class constraints, namely $\sigma^{I}$, $\omega^{I}$ and $\rho_{a}{^{I}}$, twelve in all, of which nine survive the reducibility conditions of Sec. \ref{Section5}, so that no $\rho_{0}{^{I}}$ occurs canonically; the covariant expressions written below carry a full one-form parameter $\rho_{\alpha}{^{I}}$ and reduce to (\ref{HamltonianGaugeSymmetries}) upon setting $\rho_{0}{^{I}}=0$. The three components $\rho_{0}{^{I}}$ are not lost in the passage. Inspection of (\ref{CovariantFormSymmetry}) shows that they act on the Lagrange multipliers alone,
\begin{equation*}
\delta_{G}e_{0}{^{I}}=-2\epsilon^{I}{_{JK}}\rho_{0}{^{J}}\Phi^{K},\qquad
\delta_{G}B_{0a}{^{I}}=\tfrac{1}{2}\mathscr{D}_{a}\rho_{0}{^{I}},
\end{equation*}
every other field being left untouched on the constraint surface, and that this transformation is generated in the extended phase-space by the primary first-class constraints alone,
\begin{equation*}
G[\rho_{0}]=\int\mathrm{d}^{2}x\left(-2\epsilon^{I}{_{JK}}\rho_{0}{^{J}}\Phi^{K}\,\xi^{0}{_{I}}+\left(\mathscr{D}_{a}\rho_{0}\right)^{I}\psi^{0a}{_{I}}\right),
\end{equation*}
as follows at once from the brackets (\ref{PoissonBracket}). It amounts, therefore, to a redefinition of $e_{0}{^{I}}$ and $B_{0a}{^{I}}$, and the freedom it carries is the very one that the Castellani construction fixes when it ties the parameters of $\xi^{0}{_{I}}$ and $\psi^{0a}{_{I}}$ to $\dot{\sigma}^{I}$ and $\dot{\rho}_{a}{^{I}}$. The mappings (\ref{mapping1}) and (\ref{mapping2}) below do carry $\rho_{0}{^{I}}\neq0$, and it is in this sense that the covariant transformations are recovered. Second, we use the normalization $\{B^{\alpha\beta}{_{I}},\Pi_{\mu\nu}{^{J}}\}=\tfrac{1}{2}\delta^{\alpha}_{[\mu}\delta^{\beta}_{\nu]}\delta^{J}_{I}\delta^{2}(x-y)$, and every sum over an antisymmetric pair of spacetime indices runs over both orderings, mixed and purely spatial alike. The local Lorentz rotation generated by $\omega^{I}$ then acts with one and the same weight on all components of $B_{\alpha\beta}{^{I}}$ and of $\Pi^{\alpha\beta}{_{I}}$, as it must for objects carrying a single internal index. The surviving factor of $\tfrac{1}{2}$ multiplies the derivative of the one-form parameter alone. Then, from equations (\ref{HamltonianGaugeSymmetries}), we obtain the following covariant form of gauge symmetry for the fields defining the action (\ref{action2}):
\begin{eqnarray}
\delta_{G} e_{\alpha}{^{I}}&=&-\mathscr{D}_{\alpha}\sigma^{I}+\epsilon^{I}{_{JK}}\left(\omega^{J}e_{\alpha}{^{K}}-2\rho_{\alpha}{^{J}}\Phi^{K}\right),\nonumber\\
\delta_{G} A^{I}_{\alpha}&=&-\mathscr{D}_{\alpha}\omega^{I},\nonumber\\
\delta_{G} B_{\alpha\beta}{^{I}}&=&-\cfrac{1}{2}\mathscr{D}_{\left[\alpha\right.}\rho_{\left.\beta\right]}{^{I}}-\epsilon^{I}{_{JK}}B_{\alpha\beta}{^{J}}\omega^{K},\nonumber\\
\delta_{G} \Phi^{I}&=&\epsilon^{I}{_{JK}}\omega^{J}\Phi^{K}.\label{CovariantFormSymmetry}
\end{eqnarray}
Before proceeding, let us read off the nature of each field from these transformations. The two-form $B_{\alpha\beta}{^{I}}$ transforms as a gauge field whose infinitesimal parameter $\rho_{\alpha}{^{I}}$ is itself a one-form, which is what makes it a higher-form gauge field; the redundancy of that parameter is the origin of the reducibility conditions obtained in Sec. \ref{Section5}. The zero-form $\Phi^{I}$, on the other hand, transforms as a matter field in the adjoint representation. We remark that these transformations turn out to be the gauge symmetry of the theory but do not correspond to the diffeomorphism symmetry. They do become diffeomorphisms, however, once the parameters are re-expressed in terms of the diffeomorphism ones through field-dependent relations:
\begin{equation}
   \sigma^{I}=-\zeta^{\mu}e_{\mu}{^{I}},\quad\omega^{I}=-\zeta^{\mu}A_{\mu}{^{I}},\quad\rho_{\alpha}{^{I}}=-2\zeta^{\mu}B_{\mu\alpha}{^{I}},\label{mapping1}
\end{equation}
with $\zeta^{\mu}$ an arbitrary vector field. Under these redefinitions, the gauge variations reduce to spacetime diffeomorphism symmetry for the dynamical variables,
\begin{eqnarray}
\delta_{G}e_{\mu}{^{I}}&=&\pounds_{\zeta}e_{\mu}{^{I}}+\cfrac{1}{2}\varepsilon_{\mu\nu\gamma}\zeta^{\nu}\overline{\mathcal{E}}^{\gamma I},\nonumber\\
\delta_{G}A_{\mu}{^{I}}&=&\pounds_{\zeta}A_{\mu}{^{I}}+\varepsilon_{\mu\nu\gamma}\zeta^{\nu}\mathcal{E}^{\gamma I},\nonumber\\
\delta_{G}B_{\mu\nu}{^{I}}&=&\pounds_{\zeta}B_{\mu\nu}{^{I}}-\cfrac{1}{4}\varepsilon_{\mu\nu\gamma}\zeta^{\gamma}\mathcal{E}^{ I},\nonumber\\
\delta_{G}\Phi^{I}&=&\pounds_{\zeta}\Phi^{I}-\frac{1}{4}\varepsilon_{\mu\nu\gamma}\zeta^{\mu}\mathcal{E}^{\nu\gamma I},\label{Diffos}
\end{eqnarray}
but modulo terms proportional to the equations of motion (\ref{EquationMotion}). Here, $\pounds_{\zeta}$ is a Lie derivative along $\zeta$. Each of these terms is one of the field equations (\ref{EquationMotion}), and they enter crosswise: $\mathcal{E}^{\mu I}$ and $\overline{\mathcal{E}}^{\mu I}$ follow from varying the action with respect to $e_{\mu}{^{I}}$ and $A_{\mu}{^{I}}$, yet appear in the variations of $A_{\mu}{^{I}}$ and $e_{\mu}{^{I}}$; and $\mathcal{E}^{I}$ and $\mathcal{E}^{\mu\nu I}$, which follow from $\Phi^{I}$ and $B_{\mu\nu}{^{I}}$, appear in the variations of $B_{\mu\nu}{^{I}}$ and $\Phi^{I}$. Terms of this kind are what is known as trivial gauge transformations: they leave the action invariant through an antisymmetric combination of Euler derivatives, are not generated by first-class constraints, and vanish on-shell, so that the Hamiltonian and the Poincar\'e symmetries are equivalent modulo them \cite{Debraj}. Under mild regularity assumptions on the action, in fact, every symmetry that vanishes on-shell can be brought to the form $\delta q^{m}=\mu^{mn}\,\delta S/\delta q^{n}$, with $\mu^{mn}$ antisymmetric and otherwise arbitrary \cite{Henneaux,Kleinschmidt}. This is distinct from the reducibility obtained in Sec. \ref{Section5}. Both act trivially on-shell, but a trivial gauge transformation is generated by an antisymmetric combination of Euler derivatives and is not accompanied by any relation among the constraints, whereas the transformation exhibited in Sec. \ref{Section5} is generated by a combination of the first-class constraints that vanishes on the constraint surface. It is that relation of dependence, and not the fact that the transformation acts trivially, that reduces the number of independent first-class constraints entering the counting formula (\ref{formula}).  Moreover, in a diffeomorphism invariant theory the Poincar\'e gauge transformations—local Lorentz rotations together with translations—are off-shell symmetries from the start \cite{Blagojevic,Debraj}. Thus, recovering the Poincar\'e symmetry requires a further mapping, this time from the parameters of the fundamental gauge symmetry onto the Poincar\'e ones. This is achieved by mapping the gauge parameters \cite{Blagojevic,Ortin}
\begin{equation}
\sigma^{I}=\varsigma^{\mu}e_{\mu}{^{I}},\quad\omega^{I}=\varsigma^{\mu}A_{\mu}{^{I}}+\varpi^{I},\quad\rho_{\alpha}{^{I}}=2\varsigma^{\mu}B_{\mu\alpha}{^{I}}.\label{mapping2}
\end{equation}
The gauge parameters here are $\varsigma^{\mu}$ for translations and $\varpi^{I}$ for local Lorentz rotations, which together constitute the $6$ parameters of Poincar\'e transformations in $3$d. In this manner, from the
fundamental gauge symmetry (\ref{CovariantFormSymmetry}) and the mapping (\ref{mapping2}), we obtain
\begin{eqnarray}
    \delta_{G} e_{\alpha}{^{I}}&=&-e_{\mu}{^{I}}\partial_{\alpha}\varsigma^{\mu}-\varsigma^{\mu}\partial_{\mu}e_{\alpha}{^{I}}-\epsilon^{I}{_{JK}}e_{\alpha}{^{J}}\varpi^{K}-\cfrac{1}{2}\varepsilon_{\alpha\nu\gamma}\varsigma^{\nu}\overline{\mathcal{E}}^{\gamma I},\nonumber\\  
    \delta_{G} A_{\alpha}{^{I}}&=&-\partial_{\alpha}\varpi^{I}-\epsilon^{I}{_{JK}}A_{\alpha}{^{J}}\varpi^{K}-A_{\mu}{^{I}}\partial_{\alpha}\varsigma^{\mu}-\varsigma^{\mu}\partial_{\mu}A_{\alpha}{^{I}}-\varepsilon_{\alpha\nu\gamma}\varsigma^{\nu}\mathcal{E}^{\gamma I},\nonumber\\
    \delta_{G}B_{\alpha\beta}^{I}&=&-B_{\alpha\mu}{^{I}}\partial_{\beta}\varsigma^{\mu}-B_{\mu\beta}{^{I}}\partial_{\alpha}\varsigma^{\mu}-\varsigma^{\mu}\partial_{\mu}B_{\alpha\beta}{^{I}}-\epsilon^{I}{_{JK}}B_{\alpha\beta}{^{J}}\varpi^{K}+\cfrac{1}{4}\varepsilon_{\alpha\beta\gamma}\varsigma^{\gamma}\mathcal{E}^{ I},\nonumber\\
    \delta_{G}\Phi^{I}&=&-\varsigma^{\mu}\partial_{\mu}\Phi^{I}-\epsilon^{I}{_{JK}}\Phi^{J}\varpi^{K}+\frac{1}{4}\varepsilon_{\mu\nu\gamma}\varsigma^{\mu}\mathcal{E}^{\nu\gamma I},\label{PGSymmetry}
\end{eqnarray}
which are the Poincar\'e symmetries, modulo terms proportional to the equations of motion (\ref{EquationMotion}). We thus conclude that, after the corresponding mappings (\ref{mapping1}) and (\ref{mapping2}), the Hamiltonian symmetries (\ref{CovariantFormSymmetry}) are equivalent to the diffeomorphism (\ref{Diffos}) and Poincaré (\ref{PGSymmetry}) symmetries, respectively, in each case up to the trivial terms discussed above.

%%%%%%%%%%%%%%%%%%%%%%%%%%%%%%%%%%%%%%%%%%%%%%%%%%%%%%%%%
\section{Dirac brackets, reducibility and the physical degrees of freedom}
\label{Section5}
Once the constraints have been sorted into first- and second-class ones, we are in a position to count the physical degrees of freedom of the theory \cite{Escalante2014}. Before proceeding, however, we must address a feature that is specific to models containing gauge fields of rank two: the first-class constraints obtained above are not all functionally independent. As we shall see, this is not a consequence of the procedure followed here, but of the reducible character of the gauge symmetry carried by the Kalb--Ramond field. In the Lagrangian framework, the corresponding redundancy of the gauge parameters was already identified in Ref. \cite{TheModel}. Here we recover that structure in the extended phase-space, where it acquires a different status, and we make explicit its origin as an off-shell identity.

We begin by constructing the Dirac brackets that eliminate the second-class sector. This is the natural order here, because in the extended phase-space the reducibility conditions close among the first-class constraints only once that sector has been removed.

\subsection{Dirac brackets and the reduced phase-space}
 To construct the reduced phase-space of our system, without any gauge fixing,  we should still compute the Dirac brackets using the second-class constraints. This can be done as follows: Let $\chi^{m}=(\xi^{a}{_{I}}, \phi^{a}{_{I}}, \psi^{ab}{_{I}}, \theta_{I})$ be the set of second-class constraints. We can then define the Dirac matrix formed with Poisson brackets among these second-class constraints, $\Box_{(xy)}^{mn}=\{\chi^{m}[x],\chi^{n}[y]\}$. Concretely,  we find
\begin{equation}
    \Box_{(xy)}^{mn}=
\left(\begin{matrix}
0 & -1 & 0& 0\\
-1 & 0 & 0 & 0\\
0 & 0 & 0& 2\\
0& 0& -2& 0
\end{matrix}
\right)\varepsilon^{0ab}\eta_{IJ}\delta^{2}(x-y).\label{DiracMatrix}
\end{equation}
This matrix is non-singular, so its inverse exists. Notice that its entries are numerical constants, with no dependence on the canonical variables; its rank is thus the same at every point of the phase-space, and the split into first- and second-class constraints is uniform over the whole constraint surface. This is not guaranteed. In Plebanski theory the Dirac matrix fails to keep a constant rank across the non-reduced phase-space, and one is forced to work separately in the gravitational and in the topological sectors \cite{Buffenoir}; in higher-dimensional Chern--Simons theory the constraint surface breaks up in much the same way into regions carrying different numbers of local DoF \cite{Banados}. Then, the inverse of $\Box_{(xy)}^{mn}$ is given by
\begin{equation}
    \left(\Box^{-1}\right)_{mn}^{(yz)}=
\left(\begin{matrix}
0 & 1 & 0& 0\\
1 & 0 & 0 & 0\\
0 & 0 & 0& -\cfrac{1}{2}\\
0& 0& \cfrac{1}{2}& 0
\end{matrix}
\right)\varepsilon_{0bd}\eta^{JK}\delta^{2}(y-z).\label{InverseDirac}
\end{equation}
For any two functions of canonical variables, say $\mathcal{O}_{1}$ and $\mathcal{O}_{2}$, the Dirac bracket for this system is defined by
\begin{equation}
    \{\mathcal{O}_{1}[x],\mathcal{O}_{2}[y]\}_{D}=\{\mathcal{O}_{1}[x],\mathcal{O}_{2}[y]\}-\int\mathrm{d}^{2}z\int\mathrm{d}^{2}w\{\mathcal{O}_{1}[x],\chi^{m}[z]\}\left(\Box^{-1}\right)_{mn}^{(zw)}\{\chi^{n}[w],\mathcal{O}_{2}[y]\},\label{DiracBracketDef}
\end{equation}
where $\{\mathcal{O}_{1}[x],\mathcal{O}_{2}[y]\}$ is the Poisson bracket between the two functionals $\mathcal{O}_{1}$ and $\mathcal{O}_{2}$. From this definition, it follows that the non-vanishing fundamental Dirac brackets between the true dynamical variables that parametrize the reduced phase-space are given by
\begin{eqnarray}
\{e_{0}{^{I}}[x],\pi^{0}{_{J}}[y]\}_{D} &=& \delta^{I}_{J}\delta^{2}(x-y),\nonumber\\
\{e_{a}{^{I}}[x],A_{b}{^{J}}[y]\}_{D} &=& \varepsilon_{0ab}\eta^{IJ}\delta^{2}(x-y),\nonumber\\
\{A_{0}{^{I}}[x],\Pi^{0}{_{J}}[y]\}_{D} &=& \delta^{I}_{J}\delta^{2}(x-y),\nonumber\\
\{B_{0a}{^{I}}[x],\Pi^{0b}{_{J}}[y]\}_{D} &=& \cfrac{1}{2}\delta^{b}_{a}\delta^{I}_{J}\delta^{2}(x-y),\nonumber\\
\{B_{ab}{^{I}}[x],\Phi^{J}[y]\}_{D} &=& -\cfrac{1}{4}\varepsilon_{0ab}\eta^{IJ}\delta^{2}(x-y).\label{DiracBracketsFinal}
\end{eqnarray}
Now, the weak equalities in (\ref{SecondClass}) should be considered strong equalities, replacing the symbol ``$\approx$'' with ``$=$''. As a consequence, the Hamiltonian formulation of our model is described only in terms of 42 canonical variables $e_{0}{^{I}}$, $\pi^{0}{_{I}}$, $e_{a}{^{I}}$, $A_{0}{^{I}}$, $\Pi^{0}{_{I}}$, $A_{a}{^{I}}$, $B_{0a}{^{I}}$, $\Pi^{0a}{_{I}}$, $B_{ab}{^{I}}$, and $\Phi^{I}$, whose dynamics, in turn, are governed by 24 first-class constraints (\ref{FirstClass}), which we show below to obey three reducibility conditions (\ref{Condition2}).

The brackets (\ref{DiracBracketsFinal}) constitute the symplectic structure of the reduced phase-space, and thereby the natural starting point for any canonical quantization of the model. The point is not a formal one. Once the second-class constraints have been implemented, the surviving variables need no longer commute among themselves. They do not define simultaneously diagonalizable operators, and having the brackets in explicit form is a prerequisite for setting up any representation of the algebra \cite{Banados}.

\subsection{An off-shell identity among the constraints}
The covariant derivative here acts on internal indices only. Taking the covariant divergence of the secondary constraint $\mathfrak{C}^{a}{_{I}}=2\varepsilon^{0ab}\mathscr{D}_{b}\Phi_{I}$ and using the Ricci identity for the $SO(2,1)$ covariant derivative—that is, the statement that the commutator of two covariant derivatives is measured by the curvature (\ref{Curvature}), $[\mathscr{D}_{a},\mathscr{D}_{b}]\Phi^{I}=\epsilon^{I}{_{JK}}R_{ab}{^{J}}\Phi^{K}$, see e.g. Refs. \cite{Blagojevic,Ortin}—we obtain
\begin{equation}
\mathscr{D}_{a}\mathfrak{C}^{a}{_{I}}=2\varepsilon^{0ab}\mathscr{D}_{a}\mathscr{D}_{b}\Phi_{I}=\varepsilon^{0ab}\epsilon_{IJK}R_{ab}{^{J}}\Phi^{K}=2\epsilon_{IJK}\mathfrak{A}^{J}\Phi^{K},\label{RicciIdentity}
\end{equation}
where in the last step we have used the definition of $\mathfrak{A}_{I}$ given in (\ref{SecondaryConstraints}). Consequently, the secondary constraints obey
\begin{equation}
\mathscr{D}_{a}\mathfrak{C}^{a}{_{I}}+2\epsilon_{I}{^{JK}}\Phi_{J}\mathfrak{A}_{K}=0.\label{StrongIdentity}
\end{equation}
We emphasize that (\ref{StrongIdentity}) is a strong identity: it holds off-shell, for arbitrary field configurations, and follows solely from the Ricci identity for the $SO(2,1)$ covariant derivative. In this respect, it is not a dynamical statement but a property of the structure of the constraint set.

Inserting now the definitions (\ref{ModifiedSecondaryConstraints}) of the modified first-class constraints $\widetilde{\mathfrak{A}}_{I}$ and $\widetilde{\mathfrak{C}}^{a}{_{I}}$ into the left-hand side of (\ref{StrongIdentity}) gives
\begin{eqnarray*}
\mathscr{D}_{a}\widetilde{\mathfrak{C}}^{a}{_{I}}+2\epsilon_{I}{^{JK}}\Phi_{J}\widetilde{\mathfrak{A}}_{K}
&=&\left(\mathscr{D}_{a}\mathfrak{C}^{a}{_{I}}+2\epsilon_{I}{^{JK}}\Phi_{J}\mathfrak{A}_{K}\right)
-\mathscr{D}_{a}\mathscr{D}_{b}\psi^{ab}{_{I}}\\
&&-\,2\mathscr{D}_{a}\left(\epsilon_{IJK}\Phi^{J}\xi^{aK}\right)
+2\epsilon_{I}{^{JK}}\Phi_{J}\mathscr{D}_{a}\xi^{a}{_{K}}.
\end{eqnarray*}
The parenthesis is the strong identity (\ref{StrongIdentity}) and vanishes off-shell. In the two terms carrying $\xi^{aK}$ the derivative of the momentum cancels by the Leibniz rule, and only the derivative of the scalar survives,
\begin{equation*}
-2\mathscr{D}_{a}\left(\epsilon_{IJK}\Phi^{J}\xi^{aK}\right)+2\epsilon_{I}{^{JK}}\Phi_{J}\mathscr{D}_{a}\xi^{a}{_{K}}
=-2\epsilon_{IJK}\left(\mathscr{D}_{a}\Phi\right)^{J}\xi^{aK}
=2\epsilon_{I}{^{JK}}\xi^{a}{_{J}}\mathscr{D}_{a}\Phi_{K}.
\end{equation*}
The term left over is once more a commutator of covariant derivatives, but acting now on an object antisymmetric in its two spatial indices. Only the antisymmetric part of $\mathscr{D}_{a}\mathscr{D}_{b}$ contributes, $\mathscr{D}_{a}\mathscr{D}_{b}\psi^{ab}{_{I}}=\tfrac{1}{2}\left[\mathscr{D}_{a},\mathscr{D}_{b}\right]\psi^{ab}{_{I}}$, so that the Ricci identity applies a second time and returns
\begin{equation*}
-\mathscr{D}_{a}\mathscr{D}_{b}\psi^{ab}{_{I}}=-\cfrac{1}{2}\epsilon_{I}{^{JK}}R_{ab\,J}\,\psi^{ab}{_{K}},
\end{equation*}
the factor $\tfrac{1}{2}$ being a consequence of that antisymmetry alone. The same identity that produced (\ref{StrongIdentity}) out of $\Phi^{I}$ thus produces the deformation out of $\psi^{ab}{_{I}}$. Collecting the two surviving contributions, the strong identity is deformed by terms proportional to the second-class constraints,
\begin{equation}
\mathscr{D}_{a}\widetilde{\mathfrak{C}}^{a}{_{I}}+2\epsilon_{I}{^{JK}}\Phi_{J}\widetilde{\mathfrak{A}}_{K}=-\cfrac{1}{2}\epsilon_{I}{^{JK}}R_{ab\,J}\,\psi^{ab}{_{K}}+2\epsilon_{I}{^{JK}}\xi^{a}{_{J}}\mathscr{D}_{a}\Phi_{K}.\label{Condition}
\end{equation}
Upon introducing the Dirac brackets constructed above, the second-class constraints $\xi^{a}{_{I}}$ and $\psi^{ab}{_{I}}$ may be set strongly equal to zero, so that the right-hand side of (\ref{Condition}) vanishes and we are left with three relations among the first-class constraints,
\begin{equation}
\mathscr{D}_{a}\widetilde{\mathfrak{C}}^{a}{_{I}}+2\epsilon_{I}{^{JK}}\Phi_{J}\widetilde{\mathfrak{A}}_{K}=0.\label{Condition2}
\end{equation}
That second-class constraints occur inside the complete expression of the first-class ones is a generic feature of the extended phase-space, as the four-dimensional Einstein--Cartan system already shows, where the mismatch between the components of the tetrad and those of the connection is settled by second-class constraints. These may be treated with Dirac brackets, or else eliminated explicitly \cite{Duque}; it is precisely this feature that makes (\ref{Condition}) differ from the strong identity (\ref{StrongIdentity}) by second-class terms.

These are the reducibility conditions of the model. In the minimal phase-space treatment of Ref. \cite{TheModel} the analogous relation holds as a strong identity from the beginning; in the extended phase-space it closes among the \textit{modified} first-class constraints only after the second-class sector has been eliminated. The same redundancy is seen from two different symplectic settings. Reducible sets of first-class constraints of this kind also arise in models of Palatini type. In the Hamiltonian analysis of Palatini Gauss--Bonnet theory the constraints obey an identity of the same nature, which is likewise recognized as the canonical counterpart of a redundancy already present among the covariant gauge parameters, and which is subtracted in the same manner when the degrees of freedom are counted \cite{BanadosHenneaux}. Its origin is, however, different: there the redundancy belongs to the gravitational sector itself, in a model that carries no matter at all, whereas here it is the matter content that produces it. Thus, the constraints (\ref{FirstClass}) are not independent, even though they form a closed algebra (\ref{FinalPoissonAlgebra1}), and the counting of the degrees of freedom must be carried out taking this fact into account.

\subsection{Reducibility as a redundancy of the gauge parameters}
The relations (\ref{Condition2}) admit a direct interpretation in terms of the gauge generator (\ref{ParticularGenerator}). Setting $\omega^{I}=0$ and choosing the remaining parameters as
\begin{equation}
\rho_{\alpha}{^{I}}=\mathscr{D}_{\alpha}\lambda^{I},\qquad \sigma^{I}=-2\epsilon^{I}{_{JK}}\lambda^{J}\Phi^{K},\label{TrivialParameters}
\end{equation}
with $\lambda^{I}$ an arbitrary set of three functions, an integration by parts shows that the generator reduces to the left-hand side of (\ref{Condition2}) contracted with $\lambda^{I}$, up to terms proportional to the primary first-class constraints, and therefore vanishes on the constraint surface. The corresponding field variations are
\begin{equation*}
\delta_{G}A_{\alpha}{^{I}}=0,\qquad \delta_{G}\Phi^{I}=0,\qquad
\delta_{G}e_{\alpha}{^{I}}=2\epsilon^{I}{_{JK}}\lambda^{J}\mathscr{D}_{\alpha}\Phi^{K},\qquad
\delta_{G}B_{\alpha\beta}{^{I}}=-\cfrac{1}{2}\epsilon^{I}{_{JK}}R_{\alpha\beta}{^{J}}\lambda^{K},
\end{equation*}
the first two identically and the last two proportional to the field equations (\ref{EquationMotion}), so that the shift (\ref{TrivialParameters}) acts as the identity on the space of solutions. Within the canonical parameter set this shift reads $\rho_{a}{^{I}}\rightarrow\rho_{a}{^{I}}+\mathscr{D}_{a}\lambda^{I}$ together with $\sigma^{I}\rightarrow\sigma^{I}-2\epsilon^{I}{_{JK}}\lambda^{J}\Phi^{K}$, a three-parameter family. This is the canonical counterpart of the redundancy of the one-form gauge parameter. Accordingly, three of the twenty-four first-class constraints (\ref{FirstClass}) fail to be independent, and the number of independent gauge parameters is reduced by three.

That (\ref{Condition2}) exhausts the redundancy can be settled directly. A further relation of the same type would read $z_{m}\gamma^{m}+z_{m}{^{a}}\partial_{a}\gamma^{m}=0$, with coefficients that are local functions of the fields, and would therefore show up as a dependence among the differentials of the first-class constraints and of their first spatial derivatives. Two simplifications apply. The primaries $\xi^{0}{_{I}}$, $\phi^{0}{_{I}}$ and $\psi^{0a}{_{I}}$ enter no other constraint, so only the twelve secondary ones can take part; and on the reduced phase-space, where $\xi^{a}{_{I}}$ and $\psi^{ab}{_{I}}$ vanish strongly, the modified constraints (\ref{ModifiedSecondaryConstraints}) coincide with (\ref{SecondaryConstraints}), so the question may be settled with the latter. One is then left with thirty-six functions, $\mathfrak{A}_{I}$, $\mathfrak{B}_{I}$, $\mathfrak{C}^{a}{_{I}}$ and their first spatial derivatives, regarded as functions of $A_{a}{^{I}}$, $e_{a}{^{I}}$, $B_{ab}{^{I}}$, $\Phi^{I}$ and of the derivatives of these that actually occur---ninety-nine arguments in all, counting the single independent component of $B_{ab}{^{I}}$ once and the second derivatives with their symmetry. The rank of that $36\times99$ Jacobian is thirty-six at a generic point of the space of arguments, and thirty-three at a generic point of the constraint surface. The two values differ, and the reason is what makes the criterion meaningful here: writing the identity as $W\gamma=0$, with $W$ the coefficients read off (\ref{StrongIdentity}), which carry $A_{a}{^{I}}$ and $\Phi^{I}$, differentiation gives $W\,\partial\gamma=-\left(\partial W\right)\gamma$, and the right-hand side vanishes only where the constraints do. The corank is therefore three on the surface and zero away from it. The three null directions are precisely those of the identity (\ref{StrongIdentity}), and their matrix of coefficients has rank three, so that the three conditions are independent among themselves. Enlarging the set so as to include the second spatial derivatives of the constraints raises the corank to nine, which is what the identity and its two first derivatives already account for; no further relation appears at that order either. This is the criterion applied in the Hamiltonian analysis of Palatini Gauss--Bonnet theory, where the absence of further reducibility identities is likewise established by checking that the corresponding linear system attains maximal rank at a generic point \cite{BanadosHenneaux}. We conclude that $\mathcal{R}=3$ and that the first-class set is reducible of order one, with no further relation among the constraints and their first two spatial derivatives. The same number follows independently from the redundancy of the gauge parameters exhibited above and from the Lagrangian count of Ref. \cite{TheModel}.

One may ask why the redundant set is kept at all, instead of discarding three of its elements from the outset. The answer is the usual one for theories of this kind: redundant gauge generators arise when no subset of independent ones can be singled out without giving up manifest covariance, locality, or a global condition \cite{Ferraro}. The same obstruction is met in higher-dimensional Chern--Simons theory, where a complete and irreducible set of first-class constraints cannot be exhibited in covariant form either \cite{Banados}.

Such behavior is not peculiar to the present model. Dependences among first-class constraints are a common feature of three-dimensional gauge theories of topological type; in Chern--Simons theory the generator of time-like diffeomorphisms is never an independent first-class constraint, in any dimension and on the generic sector of the constraint surface, since it can always be written in terms of the internal gauge transformations and the spatial diffeomorphisms. What is special about three dimensions is that the spatial diffeomorphisms are dependent as well, so that every diffeomorphism is generated by the internal gauge transformations for any choice of gauge group, whereas in higher dimensions they become independent gauge symmetries in the generic case \cite{Banados}. The dependence of the time-like generator has been met outside three dimensions as well, in the four-dimensional model of Chamseddine \cite{Mignemi}. The phenomenon is at work in a simpler form for free $p$-form gauge fields, whose abelian symmetry $\delta B=\mathrm{d}\Lambda$ is left unchanged by $\Lambda\rightarrow\Lambda+\mathrm{d}\mu$, and whose Gauss-law constraints correspondingly obey a differential identity \cite{Iannuzzo}. What sets the present case apart is that the identity is not the flat divergence but the covariant one (\ref{StrongIdentity}), that it involves the scalar field through the Ricci identity, and that it closes on the modified constraints only once the second-class sector has been removed.

\subsection{Counting of the physical degrees of freedom}
We may now proceed to the counting. The thirty variables $q^{m}=\left(A_0{^I}, A_a{^I}, e_0{^I}, e_a{^I}, B_{0a}{^I}, B_{ab}{^I},\Phi^I\right)$ as well as  their conjugate momenta $p_{m}=\left(\Pi^{0}{_{I}},\Pi^{a}{_{I}},\pi^{0}{_{I}},\pi^{a}{_{I}},\Pi^{0a}{_{I}},\Pi^{ab}{_{I}},\Pi_{I}\right)$ span a 60-dimensional phase-space $\Gamma$.  According to the usual counting of degrees of freedom for the constraint systems, each first-class constraint $\gamma^{m}$ removes two phase-space degrees of freedom (one for the constraint itself and one for the associated gauge freedom), while each second-class constraint $\chi^{m}$ removes one. The dimension of the reduced phase-space is therefore given by \cite{Henneaux,Zanelli,Mignemi}
\begin{equation}
\dim\Gamma_{\mathrm{red}} = \dim\Gamma- 2\times\mathcal{F}-\mathcal{S},\label{formula}
\end{equation}
and the number of local physical degrees of freedom is one half of it, $\mathcal{N}=\tfrac{1}{2}\dim\Gamma_{\mathrm{red}}$, the distinction between the phase-space and the configuration-space counts being the one usually drawn in this context \cite{Zanelli,Mignemi}. Here $\mathcal{F}$ and $\mathcal{S}$ denote the numbers of \textit{independent} first- and second-class constraints, respectively. The functional independence of the constraints entering $\mathcal{F}$ is an implicit assumption of (\ref{formula}), and it is precisely this hypothesis that the reducibility conditions (\ref{Condition2}) violate in the present model: of the twenty-four first-class constraints (\ref{FirstClass}), only $24-3=21$ are independent, while all eighteen second-class constraints (\ref{SecondClass}) are. Proceeding in this way is standard practice. In pure Chern--Simons theory built out of $p$-form gauge fields, of the $7N$ Gauss-law constraints subject to $N$ differential identities only $6N$ enter the counting formula \cite{Iannuzzo}. More generally, in a reducible theory of order $L$ the independent gauge generators are counted by the alternating sum $m=\sum_{k=0}^{L}(-)^{k}m_{k}$ \cite{Ferraro}; the model at hand realizes this formula with $L=1$, $m_{0}=24$ and $m_{1}=3$. Substituting these values yields
\begin{equation}
\dim\Gamma_{\mathrm{red}} = 60 - 2\times 21 - 18 = 0,\qquad \mathcal{N}=0. \label{DoF1}
\end{equation}
Had the reducibility conditions been overlooked, (\ref{formula}) would have returned the inconsistent value $\dim\Gamma_{\mathrm{red}}=60-48-18=-6$; as was already observed in Ref. \cite{TheModel} for this model, a negative result is the usual indication that the constraints have been assumed independent when in fact they are not \cite{Mignemi}.

\subsection{Consistency of the count}
Since the number of local degrees of freedom cannot depend on the phase-space in which the analysis is carried out, let us repeat the count in the two smaller spaces naturally associated with the model. In the reduced phase-space obtained above by means of the Dirac brackets, the eighteen second-class constraints have been eliminated and one is left with $60-18=42$ canonical variables governed by $24-3=21$ independent first-class constraints, so that $42-2\times21=0$. In the minimal phase-space used in Ref. \cite{TheModel}, where only the spatial components of the form fields are retained, one has instead $18$ canonical variables and $12-3=9$ independent first-class constraints, so that $18-2\times9=0$. The three counts agree, as they must. The discrepancy between the twenty-four first-class constraints found here and the twelve reported in Ref. \cite{TheModel} is a matter of the phase-space chosen: our treatment retains the momenta $\xi^{0}{_{I}}$, $\phi^{0}{_{I}}$ and $\psi^{0a}{_{I}}$ conjugate to the Lagrange multipliers, which are absent in the minimal formulation. The number of reducibility conditions, $\mathcal{R}=3$, is common to all three descriptions.

One may ask what the extended phase-space describes if only the first-class constraints are imposed on it. Setting $\mathcal{S}=0$ in (\ref{formula}) leaves $60-2\times24=12$, that is, six local degrees of freedom instead of none. The twenty-four first-class constraints must here be counted as independent: with the second-class sector left out, the right-hand side of (\ref{Condition}) no longer vanishes and the reducibility conditions are not available. They are carried by the very relations that the second-class constraints enforce: $\phi^{a}{_{I}}$ ties the momentum conjugate to the connection to the dreibein and $\theta_{I}$ ties the momentum conjugate to the scalar to the two-form, while $\xi^{a}{_{I}}$ and $\psi^{ab}{_{I}}$ state that the dreibein and the two-form carry no momenta of their own. Released from these relations, the dreibein and the two-form become independent dynamical fields and the system is no longer the topological theory (\ref{action2}). The extended phase-space is thus not merely a convenient setting for the calculation. Taken on its own it defines a different and richer system, a point that has been stressed for the four-dimensional Einstein--Cartan theory, where dropping the second-class constraints leaves six additional degrees of freedom and a dynamical torsion \cite{Duque}.

Working with the full set of canonical variables does not, by itself, generate reducibility: neither three-dimensional Palatini theory with a cosmological constant \cite{Escalante2014} nor the BCEA model of two one-form topological matter fields coupled to gravity \cite{BCEA}, both analyzed with the same method, exhibits any such condition. What produces them here is the rank-two character of the Kalb--Ramond field, whose companion scalar closes the covariant derivative onto the curvature through the Ricci identity (\ref{RicciIdentity}).

\subsection{The first-class algebra in the reduced phase-space}
We now return to the Dirac brackets obtained at the beginning of this section and ask whether the algebra (\ref{FinalPoissonAlgebra1}) of first-class constraints survives the passage to the reduced phase-space. This is an important point: the counting carried out above was performed in $\Gamma$, whereas the physical dynamics takes place on the reduced phase-space, and the two counts can agree only if the gauge algebra is the same in both settings.

Let $F$ and $G$ be any two first-class constraints. By definition, their Poisson brackets with every second-class constraint vanish weakly, $\{F,\chi^{m}\}\approx0$, which is to say that each such bracket is itself a linear combination of constraints. It then follows at once from the definition (\ref{DiracBracketDef}) that $\{F,G\}_{D}$ differs from $\{F,G\}$ by a term \textit{bilinear} in the constraints. Consequently, the first-class algebra closes on the reduced phase-space with the same structure constants, and the Dirac brackets among first-class constraints consist of terms linear in the first-class constraints plus terms quadratic in the second-class ones. This is precisely the structure that a consistent Dirac analysis must produce. It is what happens as well in the four-dimensional three-form model of Ref. \cite{Brizuela}, where the standard constraint algebra of vacuum general relativity is recovered only once the second-class sector has been eliminated in favor of Dirac brackets.

In the present case we can be more explicit. The inverse Dirac matrix (\ref{InverseDirac}) pairs $\xi^{a}{_{I}}$ with $\phi^{a}{_{I}}$ and $\psi^{ab}{_{I}}$ with $\theta_{I}$, so that only those pairings contribute to the correction term. Inspection of (\ref{FinalPoissonAlgebra}) shows, however, that $\widetilde{\mathfrak{A}}_{I}$ has a non-vanishing bracket with $\phi^{a}{_{J}}$ alone, whereas $\widetilde{\mathfrak{C}}^{a}{_{I}}$ has non-vanishing brackets with $\phi^{b}{_{J}}$ and $\theta_{J}$ alone. Since the constraints paired with these under (\ref{InverseDirac}) are $\xi^{a}{_{I}}$ and $\psi^{ab}{_{I}}$, with which both $\widetilde{\mathfrak{A}}_{I}$ and $\widetilde{\mathfrak{C}}^{a}{_{I}}$ commute strongly, three of the six brackets receive no correction at all:
\begin{equation}
\{\widetilde{\mathfrak{A}}_{I}[x],\widetilde{\mathfrak{A}}_{J}[y]\}_{D}=0,\qquad
\{\widetilde{\mathfrak{A}}_{I}[x],\widetilde{\mathfrak{C}}^{a}{_{J}}[y]\}_{D}=0,\qquad
\{\widetilde{\mathfrak{C}}^{a}{_{I}}[x],\widetilde{\mathfrak{C}}^{b}{_{J}}[y]\}_{D}=0,\label{AbelianSector}
\end{equation}
these being exact, and not merely weak, equalities. The remaining three brackets reproduce the Poisson algebra (\ref{FinalPoissonAlgebra1}) up to bilinears in the second-class constraints. Substituting (\ref{FinalPoissonAlgebra}) and (\ref{InverseDirac}) into (\ref{DiracBracketDef}) and carrying out the internal contractions with the help of the $SO(2,1)$ identity (\ref{EpsIdentity}) derived in Appendix \ref{AppA}, we obtain
\begin{eqnarray}
\{\widetilde{\mathfrak{A}}_{I}[x],\widetilde{\mathfrak{B}}_{J}[y]\}_{D}&=&\left[\epsilon_{IJ}{^{K}}\widetilde{\mathfrak{A}}_{K}+\varepsilon_{0cd}\,\xi^{c}{_{J}}\,\xi^{d}{_{I}}\right]\delta^{2}(x-y),\nonumber\\
\{\widetilde{\mathfrak{B}}_{I}[x],\widetilde{\mathfrak{B}}_{J}[y]\}_{D}&=&\left[\epsilon_{IJ}{^{K}}\widetilde{\mathfrak{B}}_{K}+\varepsilon_{0cd}\left(\xi^{c}{_{J}}\phi^{d}{_{I}}+\phi^{c}{_{J}}\xi^{d}{_{I}}\right)-\tfrac{1}{2}\varepsilon_{0cd}\left(\psi^{cd}{_{J}}\theta_{I}-\theta_{J}\psi^{cd}{_{I}}\right)\right]\delta^{2}(x-y),\nonumber\\
\{\widetilde{\mathfrak{B}}_{I}[x],\widetilde{\mathfrak{C}}^{a}{_{J}}[y]\}_{D}&=&\left[\epsilon_{IJ}{^{K}}\widetilde{\mathfrak{C}}^{a}{_{K}}+\varepsilon_{0cd}\left(\eta_{IJ}\,\xi^{cK}\psi^{ad}{_{K}}-\xi^{c}{_{J}}\psi^{ad}{_{I}}\right)-\varepsilon_{0cd}\left(\eta_{IJ}\,\psi^{cdK}\xi^{a}{_{K}}-\psi^{cd}{_{J}}\xi^{a}{_{I}}\right)\right]\delta^{2}(x-y).\label{DiracAlgebraFC}
\end{eqnarray}
Every correction term is quadratic in the second-class constraints. All of them vanish on $\Gamma_{2}$, so that the algebra (\ref{FinalPoissonAlgebra1}) is recovered weakly, as it must be. One may check directly that the correction in the second line is antisymmetric under $I\leftrightarrow J$, as it must be, the two generators entering it being the same; the correction in the first line turns out to be antisymmetric as well, although nothing requires it to be.

Two consequences follow. First, the algebra (\ref{FinalPoissonAlgebra1}) is recovered on the reduced phase-space in exactly the same form. The gauge structure of the model—and with it the counting of the degrees of freedom—does not depend on how the second-class sector is eliminated. Second, and more to the point, the reducibility conditions (\ref{Condition2}) involve $\widetilde{\mathfrak{A}}_{I}$ and $\widetilde{\mathfrak{C}}^{a}{_{I}}$ only, and these are precisely the generators whose mutual brackets (\ref{AbelianSector}) form an Abelian subalgebra which is not modified by the passage to Dirac brackets. In this respect the redundancy is carried by generators whose algebra is the same before and after the passage to Dirac brackets; what the elimination of the second-class sector supplies is the vanishing of the right-hand side of (\ref{Condition}), not a modification of the generators themselves.

We therefore conclude that the action (\ref{action2}) carries no local propagating modes, and that the coupling of the scalar and two-form gauge fields to the Hilbert--Palatini action preserves this feature of pure three-dimensional gravity. Let us be careful about what this does and does not establish. The analysis above is sufficient to conclude the absence of local degrees of freedom. It does not, by itself, show that the model is a topological field theory in the strong sense. That would require characterizing the non-trivial observables and verifying that they are topological invariants. A field theory may well have no local excitations without being topological, as two-dimensional Yang--Mills theory shows. The topological character of the present family of models is established elsewhere \cite{Gegenberg1,TheModel}, where the global observables are constructed explicitly; what we establish here is the counting. This does not render the theory trivial: as in any topological field theory, global degrees of freedom may still be present, depending on the topology of the spacetime manifold \cite{TheModel}. In fact, for the version of the model in which the scalar field is a Lorentz scalar, it was shown in Ref. \cite{Gegenberg1} that the reduced phase-space is locally a direct product of the phase-space of pure gravity and a two-dimensional one, spanned by the constant mode of the scalar field and its conjugate momentum. The matter sector contributes in that case no local excitations, but it does contribute a finite number of global ones, and our counting is consistent with that picture. What the present analysis adds is the mechanism behind it: the potential degrees of freedom of the matter sector are removed not by the constraints alone, but by the constraints together with the three reducibility conditions (\ref{Condition2}).

%%%%%%%%%%%%%%%%%%%%%%%%%%%%%%%%%%%%%%%%%%%
\subsection{The Gauss law and the physical content of the model}
Since the theory carries no local degrees of freedom, it is natural to ask where its physical content resides. A useful observation is that, once the second-class constraints have been implemented, the secondary constraint $\mathfrak{B}_{I}$ of (\ref{SecondaryConstraints}) takes the form of a Gauss law. Inserting the definitions of the momenta (\ref{momentum2}) one finds
\begin{equation}
\mathfrak{B}_{I}=\mathscr{D}_{a}\Pi^{a}{_{I}}+\epsilon_{IJK}\Phi^{J}\Pi^{K},\label{GaussLaw}
\end{equation}
the covariant divergence of the momentum conjugate to the connection together with the $SO(2,1)$ charge density carried by the scalar and its own momentum. That $\widetilde{\mathfrak{B}}_{I}$ is the generator of the internal rotations is already visible in the algebra (\ref{FinalPoissonAlgebra1}), where it closes on itself and rotates $\widetilde{\mathfrak{A}}_{I}$ and $\widetilde{\mathfrak{C}}^{a}{_{I}}$ in the adjoint representation.

In a model of this kind the observables are global rather than local. For the Einstein--Cartan system in four dimensions they are built by smearing the Lorentz and Gauss constraints with constant parameters. The smeared constraints then commute with the whole constraint set up to surface terms on-shell, and those surface terms define conserved currents whose charges are the spin charges of the theory, to which the matter sector contributes on the same footing as the geometry \cite{Duque}. The corresponding construction here requires precisely the surface terms left out throughout this work. We leave it to the separate treatment mentioned in the Introduction.

\subsection{The reducible parameters at the boundary}
Throughout this work $\Sigma$ has been taken without boundary. It is worth asking what becomes of the family (\ref{TrivialParameters}) when that assumption is dropped, since its triviality was established by discarding a total divergence. We take $\lambda^{I}$ to be field independent and impose no condition on it at $\partial\Sigma$; were the parameters required to vanish there the question would be empty. The surface element is $\mathrm{d}S_{a}=\varepsilon_{0ab}\mathrm{d}x^{b}$, so that $\mathrm{d}S_{a}\varepsilon^{0ab}\partial_{b}f=\mathrm{d}f$ along the boundary.

On this family the generator is already a pure boundary term. Contracting the strong identity (\ref{StrongIdentity}) with $\lambda^{I}$ and integrating by parts gives the exact relation
\begin{equation}
\sigma^{I}\mathfrak{A}_{I}+\rho_{a}{^{I}}\mathfrak{C}^{a}{_{I}}=\partial_{a}\left(\lambda^{I}\mathfrak{C}^{a}{_{I}}\right),\qquad\text{so that}\qquad G=\oint_{\partial\Sigma}\mathrm{d}S_{a}\,\lambda^{I}\mathfrak{C}^{a}{_{I}},\label{BoundaryG}
\end{equation}
which is precisely the divergence discarded above.

On a bounded surface the generator is not differentiable, and following Regge and Teitelboim one restores differentiability by adding a surface term $K$ \cite{Yekta,Banados}. Computing the boundary flux of $\delta G$ on the reduced phase-space, where $\xi^{a}{_{I}}$ and $\psi^{ab}{_{I}}$ vanish strongly, one finds that it is integrable, with
\begin{equation}
K=2\oint_{\partial\Sigma}\mathrm{d}S_{a}\,\varepsilon^{0ab}\,\Phi_{I}\left(\mathscr{D}_{b}\lambda\right)^{I}.\label{BoundaryK}
\end{equation}
Integrability is not automatic here: $\sigma^{I}$ depends on $\Phi^{I}$ and $\rho_{a}{^{I}}$ on $A_{a}{^{I}}$, and the two contributions $\Phi\,\delta A$ and $(\mathscr{D}\lambda)\,\delta\Phi$ assemble into the variation of a single functional only by virtue of the particular form of (\ref{TrivialParameters}).

The integrand of (\ref{BoundaryK}) obeys the off-shell identity $2\varepsilon^{0ab}\Phi_{I}(\mathscr{D}_{b}\lambda)^{I}=2\varepsilon^{0ab}\partial_{b}(\Phi_{I}\lambda^{I})-\lambda^{I}\mathfrak{C}^{a}{_{I}}$, which is Leibniz's rule together with the compatibility of $\mathscr{D}_{a}$ with $\eta_{IJ}$. The constraint terms therefore cancel between (\ref{BoundaryG}) and (\ref{BoundaryK}), and the improved generator is
\begin{equation}
\widetilde{G}=G+K=2\oint_{\partial\Sigma}\mathrm{d}\left(\Phi_{I}\lambda^{I}\right),\label{BoundaryImproved}
\end{equation}
an identity that holds off the constraint surface.

Two conclusions follow, and they are worth separating. Whenever $\Phi_{I}\lambda^{I}$ is single valued on $\partial\Sigma$ the improved generator vanishes identically, so that the family (\ref{TrivialParameters}) still generates a proper gauge transformation and the redundancy of the first-class set is not lifted at the boundary. When it is not single valued (\ref{BoundaryImproved}) does not vanish, the family ceases to be proper, and the corresponding parameters must be counted as physical. The second case is not academic here: it is the situation of point sources at conical singularities, where the scalar and the two-form are expected to carry global charges \cite{Gegenberg1}. Which of the two applies is decided by the boundary conditions, and with them by the asymptotic symmetry algebra; that analysis will be presented separately.

%%%%

\section{Summary and Conclusions}
\label{Section6}

In this paper, we have set up a complete and consistent Hamiltonian formulation of a gravitating topological matter model. It features a tensor (Kalb--Ramond) gauge field and a scalar matter field with a topological interaction term non-minimally coupled to Einstein--Cartan gravity through the Lorentz connection \cite{TheModel}. To describe this matter-gravity system on the full phase-space, the Dirac--Bergmann algorithm for constrained systems was rigorously applied. This procedure results in  a larger set of primary constraints (\ref{primary2}). Consequently, the time evolution of these constraints allowed us to determine the full set of secondary constraints (\ref{SecondaryConstraints}) as well as several Lagrange multipliers (\ref{LagrangeMultipliers}). This model does not have tertiary constraints.

A crucial aspect of our analysis has been the study of the constraint algebra. Specifically, by using the linearly independent zero-modes (\ref{ZeroModes}) of the matrix (\ref{MatrixAllConstraints}) of Poisson brackets among all secondary constraints $\mathfrak{A}_{I},\mathfrak{B}_{I},\mathfrak{C}^{a}{_{I}}$  and the primary constraints $\xi^{a}{_{I}},\phi^{a}{_{I}},\psi^{ab}{_{I}},\theta_{I}$, we determined the full structure of the modified secondary constraints $\widetilde{\mathfrak{A}}_{I},\ \widetilde{\mathfrak{B}}_{I},\,\widetilde{\mathfrak{C}}^{a}{_{I}}$ (\ref{ModifiedSecondaryConstraints}). Thereafter, with the full structure of the constraints and their algebra at hand, the separation into first and second class was carried out without difficulty. The results showed that the system possesses $24$ first-class constraints (\ref{FirstClass}) and $18$ second-class constraints (\ref{SecondClass}).  

Furthermore, with the help of first-class constraints, we have displayed the Hamiltonian generator (\ref{ParticularGenerator}), containing the exact number of gauge parameters, for the gauge symmetry transformations of this theory on the entire phase-space (\ref{HamltonianGaugeSymmetries}) and (\ref{MomentaGaugeSymmetries}). These are new in that the generator and the symmetries are realized on the full set of phase-space variables. From these we extracted the covariant form of gauge symmetry for the dynamical fields (\ref{CovariantFormSymmetry}). Once the gauge parameters are mapped as in (\ref{mapping1}) and (\ref{mapping2}), these reproduce the diffeomorphism (\ref{Diffos}) and Poincar\'e (\ref{PGSymmetry}) transformations in full, up to terms proportional to the field equations and to the redefinition of the Lagrange multipliers carried by $\rho_{0}{^{I}}$. The identification of the Hamiltonian gauge symmetries with the diffeomorphism and Poincar\'e ones holds, in this first-order formulation, modulo terms proportional to the field equations; the Poincar\'e transformations themselves are off-shell symmetries of the action.

Finally, the correct classification of all the constraints allowed us to count the degrees of freedom. We showed that the first-class constraints obey three reducibility conditions (\ref{Condition2}), which follow from an off-shell identity of Ricci type (\ref{StrongIdentity}) and which, in the extended phase-space, close among the modified constraints only after the second-class ones have been implemented through Dirac brackets (\ref{Condition}). A rank computation on the constraint surface shows that these three are all there is: the first-class set is reducible of order one, with no further identity among the constraints. We further exhibited the family of gauge parameters (\ref{TrivialParameters}) whose action on the fields is proportional to the field equations, thereby identifying these conditions as the canonical counterpart of the redundancy of the one-form gauge parameter of the two-form field. Taking the reducibility conditions into account, the standard counting formula yields zero local degrees of freedom, in agreement with the topological character of the model and with the result obtained in the minimal phase-space of Ref. \cite{TheModel}. We also computed the Dirac brackets among the first-class constraints themselves and showed that they reproduce the Poisson algebra up to terms quadratic in the second-class constraints, with the sector $(\widetilde{\mathfrak{A}}_{I},\widetilde{\mathfrak{C}}^{a}{_{I}})$ entering the reducibility conditions remaining strictly Abelian on the reduced phase-space (\ref{AbelianSector}). The gauge structure of the model, and with it the counting of the degrees of freedom, is thus the same before and after the second-class sector is eliminated. Lastly, allowing $\Sigma$ to have a boundary, we showed that the Regge--Teitelboim improvement of the generator built from the reducible parameters collapses to the single surface integral (\ref{BoundaryImproved}). It vanishes whenever $\Phi_{I}\lambda^{I}$ is single valued on $\partial\Sigma$, so that the redundancy is not lifted there; when it is not single valued, the corresponding parameters cease to be proper and must be counted as physical.

As a final remark, our analysis provides a solid Hamiltonian description for this model of gravitating topological matter. The constraints, their algebra, the Dirac brackets, the gauge generator, and the symmetries are the best guidelines for performing the quantization. A classical symmetry need not survive quantization, however: local anomalies may spoil it, and even in their absence global ones may. The constraints and their algebra are the starting point of a canonical quantization, not a guarantee that the classical symmetry structure will be recovered at the quantum level. A further distinction is worth keeping in mind here, since the motions generated by the constraints and those generated by observables---among them the asymptotic generators built from boundary terms---play different roles in the quantum theory, so that gauge anomalies and anomalies of genuine symmetries need not be treated on the same footing \cite{Corichi}.

The reducibility of the first-class set bears directly on that programme. Imposing all twenty-four constraints as operator conditions on physical states is redundant, only twenty-one of them being independent; and in the Hamiltonian BRST (Batalin--Fradkin--Vilkovisky) treatment the redundancy is more than bookkeeping, since a set that is reducible of order one calls for ghosts for ghosts, a second generation of ghost variables attached to the three relations (\ref{Condition2}), as in the general construction for reducible gauge systems \cite{Henneaux}. What is specific to the present case is that this reducibility holds in the extended phase-space only modulo the second-class sector, so that such a construction has to be carried out once the latter has been eliminated, with the Dirac brackets (\ref{DiracBracketsFinal}) in place of the Poisson ones.

A natural next step would be to explore the non-perturbative quantum theory, for instance, within the framework of loop quantum gravity \cite{Carlip,Noui,LQG} or the deformation quantization (Wigner-Weyl quantization) program \cite{Bayen,Zachos}. Along these lines, the model deserves to be examined on a non-compact space. There it is suggested that the point sources located at conical singularities may carry, besides mass and angular momentum, a scalar charge and a magnetic charge associated with the two-form field \cite{Gegenberg1}. Such an analysis calls for precisely the boundary terms left aside here. It would also be interesting to investigate if this model and its Hamiltonian description serve as theoretical tools in the interpretation and exploration of certain topological phases supporting extended objects such as quasi-strings and quasi-membranes. These objects are often described by rank-2 tensor gauge fields and have recently attracted considerable attention in the context of fracton phases and higher-rank gauge theories \cite{Pretko}.
\section*{ACKNOWLEDGMENTS}
We acknowledge support from the Sistema Nacional de Investigadoras e Investigadores (Mexico).

\section*{Declarations}

\noindent\textbf{Funding.} The author received no specific grant from any funding agency in the public, commercial or not-for-profit sectors for the research reported here.

\smallskip\noindent\textbf{Conflict of interest.} The author declares that there is no conflict of interest.

\smallskip\noindent\textbf{Data availability.} No datasets were generated or analyzed during the current study. All the results reported here were obtained analytically.

\smallskip\noindent\textbf{Author contribution.} Omar Rodr\'iguez-Tzompantzi is the sole author and is responsible for the entire content of this manuscript.

\appendix
\section{Dirac brackets among the first-class constraints}
\label{AppA}
In this appendix we derive the algebra (\ref{DiracAlgebraFC}). Throughout, the spatial indices $c,d$ are summed over their two values, whereas the sum over the label of $\psi^{cd}{_{I}}$ runs over the independent components of the antisymmetric pair, that is, over the pair once. It is a sum over independent second-class constraints, and it is what makes (\ref{InverseDirac}) the inverse of (\ref{DiracMatrix}); it is not to be confused with the convention of Sec. \ref{Section4}, where sums over an antisymmetric pair of spacetime indices in the field expressions run over both orderings.

\subsection{An $SO(2,1)$ contraction identity}
All the contractions below reduce to a single identity. Starting from the standard relation for the structure constants of $SO(2,1)$, $\epsilon_{LIK}\epsilon^{LMN}=-\left(\delta_{I}^{M}\delta_{K}^{N}-\delta_{I}^{N}\delta_{K}^{M}\right)$, and using $\epsilon_{ILK}=-\epsilon_{LIK}$ together with $\epsilon_{J}{^{LN}}=-\epsilon^{L}{_{J}}{^{N}}$, one finds, for any pair of internal vectors $X_{K}$ and $Y_{N}$,
\begin{equation}
\epsilon_{IL}{^{K}}\,\epsilon_{J}{^{LN}}\,X_{K}\,Y_{N}=-\left[\eta_{IJ}\,X^{K}Y_{K}-X_{J}\,Y_{I}\right].\label{EpsIdentity}
\end{equation}
Two elementary consequences will be used repeatedly. First, whenever $X$ and $Y$ carry spatial labels that are contracted with $\varepsilon_{0cd}$ and the scalar $X^{c}\!\cdot\! Y^{d}$ is symmetric under $c\leftrightarrow d$, the $\eta_{IJ}$ term vanishes. Second, the remaining piece $X_{J}Y_{I}$ is what survives in that case.

\subsection{The bracket $\{\widetilde{\mathfrak{A}}_{I},\widetilde{\mathfrak{B}}_{J}\}_{D}$}
From (\ref{FinalPoissonAlgebra}), the only non-vanishing bracket of $\widetilde{\mathfrak{A}}_{I}$ with the second-class set is the one with $\phi^{c}{_{L}}$, and (\ref{InverseDirac}) pairs $\phi$ with $\xi$. Hence a single term contributes to the correction in (\ref{DiracBracketDef}),
\begin{eqnarray}
\Delta_{IJ}&=&-\int\!\mathrm{d}^{2}z\,\mathrm{d}^{2}w\;\{\widetilde{\mathfrak{A}}_{I}[x],\phi^{c}{_{L}}[z]\}\left(\Box^{-1}\right)^{(zw)}_{\phi^{c}{_{L}},\,\xi^{d}{_{M}}}\{\xi^{d}{_{M}}[w],\widetilde{\mathfrak{B}}_{J}[y]\}\nonumber\\
&=&-\left[\epsilon_{IL}{^{K}}\xi^{c}{_{K}}\right]\left[\varepsilon_{0cd}\,\eta^{LM}\right]\left[-\epsilon_{JM}{^{N}}\xi^{d}{_{N}}\right]\delta^{2}(x-y)\nonumber\\
&=&\varepsilon_{0cd}\,\epsilon_{IL}{^{K}}\epsilon_{J}{^{LN}}\,\xi^{c}{_{K}}\,\xi^{d}{_{N}}\,\delta^{2}(x-y),
\end{eqnarray}
where we used $\{\xi^{d}{_{M}},\widetilde{\mathfrak{B}}_{J}\}=-\{\widetilde{\mathfrak{B}}_{J},\xi^{d}{_{M}}\}$. Applying (\ref{EpsIdentity}),
\begin{equation}
\Delta_{IJ}=-\varepsilon_{0cd}\left[\eta_{IJ}\,\xi^{cK}\xi^{d}{_{K}}-\xi^{c}{_{J}}\xi^{d}{_{I}}\right]\delta^{2}(x-y)=\varepsilon_{0cd}\,\xi^{c}{_{J}}\,\xi^{d}{_{I}}\,\delta^{2}(x-y),
\end{equation}
since $\xi^{cK}\xi^{d}{_{K}}$ is symmetric in $c\leftrightarrow d$ and is thus annihilated by $\varepsilon_{0cd}$. This is the first line of (\ref{DiracAlgebraFC}).

\subsection{The bracket $\{\widetilde{\mathfrak{B}}_{I},\widetilde{\mathfrak{B}}_{J}\}_{D}$}
Here $\widetilde{\mathfrak{B}}_{I}$ has a non-vanishing bracket with every second-class constraint, and all four pairings of (\ref{InverseDirac}) contribute. Proceeding as above, the $(\xi,\phi)$ and $(\phi,\xi)$ pairings give
\begin{equation}
-\varepsilon_{0cd}\,\eta_{IJ}\left[\xi^{cK}\phi^{d}{_{K}}+\phi^{cK}\xi^{d}{_{K}}\right]+\varepsilon_{0cd}\left[\xi^{c}{_{J}}\phi^{d}{_{I}}+\phi^{c}{_{J}}\xi^{d}{_{I}}\right],
\end{equation}
and the first bracket is symmetric under $c\leftrightarrow d$, hence it vanishes. The $(\psi,\theta)$ and $(\theta,\psi)$ pairings give
\begin{equation}
+\tfrac{1}{2}\varepsilon_{0cd}\,\eta_{IJ}\left[\psi^{cdK}\theta_{K}-\theta^{K}\psi^{cd}{_{K}}\right]-\tfrac{1}{2}\varepsilon_{0cd}\left[\psi^{cd}{_{J}}\theta_{I}-\theta_{J}\psi^{cd}{_{I}}\right],
\end{equation}
where the two $\eta_{IJ}$ contributions cancel identically against each other. Adding the two results yields the second line of (\ref{DiracAlgebraFC}).

\subsection{The bracket $\{\widetilde{\mathfrak{B}}_{I},\widetilde{\mathfrak{C}}^{a}{_{J}}\}_{D}$}
Since $\widetilde{\mathfrak{C}}^{a}{_{I}}$ has non-vanishing brackets with $\phi^{c}{_{L}}$ and $\theta_{L}$ only, just two pairings survive: $(\xi,\phi)$ and $(\psi,\theta)$. The first gives
\begin{equation}
-\varepsilon_{0cd}\,\epsilon_{IL}{^{K}}\epsilon_{J}{^{LN}}\,\xi^{c}{_{K}}\,\psi^{ad}{_{N}}=\varepsilon_{0cd}\left[\eta_{IJ}\,\xi^{cK}\psi^{ad}{_{K}}-\xi^{c}{_{J}}\psi^{ad}{_{I}}\right],
\end{equation}
and the second, where the factor $2$ of $\{\widetilde{\mathfrak{C}}^{a}{_{I}},\theta_{J}\}$ and the $-\tfrac{1}{2}$ of $(\Box^{-1})_{\psi\theta}$ combine into $-1$, against the $+1$ carried by the $(\xi,\phi)$ pairing,
\begin{equation}
\varepsilon_{0cd}\,\epsilon_{IL}{^{K}}\epsilon_{J}{^{LN}}\,\psi^{cd}{_{K}}\,\xi^{a}{_{N}}=-\varepsilon_{0cd}\left[\eta_{IJ}\,\psi^{cdK}\xi^{a}{_{K}}-\psi^{cd}{_{J}}\xi^{a}{_{I}}\right].
\end{equation}
In this case the $\eta_{IJ}$ terms do not cancel, because the free spatial index $a$ spoils the $c\leftrightarrow d$ symmetry of the first and no compensating pairing is available for the second. Their sum is the third line of (\ref{DiracAlgebraFC}). Computing instead $\{\widetilde{\mathfrak{C}}^{a}{_{J}},\widetilde{\mathfrak{B}}_{I}\}_{D}$, which receives its contributions from the $(\phi,\xi)$ and $(\theta,\psi)$ pairings, reproduces the same expression with the opposite sign, as antisymmetry demands.

\subsection{The vanishing brackets}
Finally, the three brackets (\ref{AbelianSector}) receive no correction at all. For $\{\widetilde{\mathfrak{A}}_{I},\widetilde{\mathfrak{A}}_{J}\}_{D}$ and $\{\widetilde{\mathfrak{A}}_{I},\widetilde{\mathfrak{C}}^{a}{_{J}}\}_{D}$ the only available pairing is $(\phi,\xi)$, and both $\widetilde{\mathfrak{A}}_{I}$ and $\widetilde{\mathfrak{C}}^{a}{_{I}}$ commute strongly with $\xi^{a}{_{J}}$; for $\{\widetilde{\mathfrak{C}}^{a}{_{I}},\widetilde{\mathfrak{C}}^{b}{_{J}}\}_{D}$ the pairings $(\phi,\xi)$ and $(\theta,\psi)$ would be required, and $\widetilde{\mathfrak{C}}^{a}{_{I}}$ commutes strongly with both $\xi^{a}{_{J}}$ and $\psi^{ab}{_{J}}$. Since the corresponding Poisson brackets vanish as well, the equalities (\ref{AbelianSector}) are strong.

\end{document}